\documentclass[9pt,onecolumn,oneside]{article}

\usepackage{authblk}
\usepackage{xcolor}
\usepackage{verbatim}
\usepackage{graphicx}
\usepackage{amsmath}
\usepackage{url}
\usepackage{hyperref}
\usepackage[margin=1.3in]{geometry}

\begin{document}

\title{Indirect social influence and diffusion of innovations: An experimental approach}

\author[a,1]{Manuel Miranda}
\author[b,c]{Mar\'ia Pereda}
\author[c,d]{Angel S\'anchez}
\author[a,1]{Ernesto Estrada}

\affil[a]{Instituto de F\'isica Interdisciplinar y Sistemas Complejos IFISC (UIB-CSIC), 07122 Palma de Mallorca, Spain}
\affil[b]{Grupo de Investigaci\'on Ingenier\'ia de Organizaci\'on y Log\'istica (IOL), Departamento Ingenier\'ia de Organizaci\'on, Administraci\'on de empresas y Estad\'istica, Escuela T\'ecnica Superior de Ingenieros Industriales, Universidad Polit\'ecnica de Madrid, Madrid 28006, Spain}
\affil[c]{Grupo Interdisciplinar de Sistemas Complejos (GISC), Departamento de Matem\'aticas, Universidad Carlos III de Madrid, 28911 Legan\'es, Spain}
\affil[d]{Instituto de Biocomputaci\'on y F\'isica de Sistemas Complejos, Universidad de Zaragoza, Zaragoza 50018, Spain}


\maketitle

\footnotetext[1]{Corresponding authors. E-mails: \url{estrada@ifisc.uib-csic.es} ; \url{mmiranda@ifisc.uib-csic.es}}

\section*{Abstract}

A fundamental feature for understanding the diffusion of innovations through a social group is the manner in which we are influenced by our own social interactions. It is usually assumed that only direct interactions, those that form our social network, determine the dynamics of adopting innovations. Here, we put this assumption to the test by experimentally and theoretically studying the role of direct and indirect influences in the adoption of innovations. We perform experiments specifically designed to
capture the influence that an individual receives from their direct social ties as well as from those socially close to them, as a function of the separation they have in their social network. The results of 21 experimental sessions with more than 590 participants show that the rate of adoption of an innovation is significantly influenced not only by our nearest neighbors but also by the second and third levels of influences an adopter has. Using a mathematical model that accounts for both direct and indirect interactions in a network, we fit the experimental results and determine the way in which influences decay with social distance. The results indicate that the strength of peer pressure on an adopter coming from its second and third circles of influence is approximately 2/3 and 1/3, respectively, relative to their closest neighbors. Our results strongly suggest that innovation adoption is a complex process in which an individual feels significant pressure not only from their direct ties but also by those socially close to them.

\section{Introduction}

In a world of traditions, an innovation is an idea, practice or object
that is perceived as new by an individual \cite[p.12]{rogers-2003}. The adoption of an innovation is not a trivial process, sometimes requiring a lengthy period of time. Once an innovation of interest arises, individuals and organizations often need to accelerate their adoption
\cite{aral-2012,ge-2014,goncalves-2012,zino-2022} for reasons that go from public health policies \cite{chami-2017,centola-2010} to  marketing ones \cite{bakshy-2012A,riedl-2018}. Therefore, understanding the diffusion of innovations, i.e., how innovations
are ``communicated through certain channels over time among the members of a social group'' \cite[p.5]{rogers-2003} is of vital importance in many areas of social sciences research \cite{rolfe-2012,diazjose-2015,bakshy-2012A,aral-2011B,bapna-2015,abella-2023}. Starting with the seminal book of Rogers, first published in the 1960's \cite{rogers-2003},  several works have attempted to find mechanisms for accelerating innovation diffusion, either from the theoretical or the experimental point of view \cite{ge-2014,guidolin-2023,aral-2011C,estrada-2013,valente-1999}. 

A specific, but quite generic case of innovation diffusion relies on communication channels formed by interpersonal communication, by means of which an individual persuades
others to accept a new idea \cite{chikouche-2018,hamblin-1975,zhang-2016}. In principle, this kind of communication channel may be understood as a social network in which pairs of individuals are
connected if they share an interpersonal communication \cite{ge-2014,hamblin-1979},
be it face-to-face exchange or via e-communication, such as
email or social media \cite{guille-2013}. However, as far as diffusion of innovations is concerned, the communication structure may be
so complex that it goes beyond the interpersonal channels of communication
recorded in the social network structure \cite{bartal-2019,aral-2014}. Indeed, as Rogers put it, ``even the members of the system may not understand the communication
structure of which they are part.'' \cite[p.337]{rogers-2003}. One of
the reasons for which the network of interpersonal ties does not capture
the totality of the communication channels is that individuals can
learn from observation of other people's behavior by means of non-verbal exchange of information. This mimicry of other's behaviors
conforms a phenomenon known as ``social'' or ``observational''
learning \cite{bandura-2000,salganik-2009,hamblin-2021}. In fact, even knowledge
about certain statistics can act as a trigger of observational learning.
One example is provided by Åberg \cite{aberg-2011} who cited the case of local demographic
as an important influence on the risk of divorce. That is, knowing
that some people not different from me have a certain behavior makes
me copy them \cite{grujic-2020,christakis-2012, pitcher-1978,fowler-2008B}. Several similar examples are given in \cite{hedstrom-2011}.

A fundamental research problem is then how to account for the combination
of interpersonal channels and observational learning into a unified
network representation of communication channels for the diffusion
of innovations. Here we take advantage of the seminal ideas of Granovetter
\cite{granovetter-1973} who assumes that all decision makers are influenced
by everyone else in an ``all see all'' network. However, not everyone
is equally considered in these interactions \cite{festinger-1954} as people
often respond most to the behaviors of those similar to them in terms of common beliefs, education, socioeconomic
status, and so forth \cite{mcpherson-2001}. Thus, Granovetter \cite{granovetter-1978}
proposed that there is a range of ties with different strengths, where
``the strength of a tie is a (probably linear) combination of the
amount of time, emotional intensity, the intimacy (mutual confiding)
and the reciprocal services that characterize the tie''. The problem
is again how to quantify these strengths of ties. In this work, we
use the ideas of Simmel about social distance \cite{simmel-1950}. According to Simmel, social distance measures the nearness or intimacy that an individual or group feels towards another individual. Therefore, it is natural to associate the concept of social distance to that of communication proximity, due to their natural equivalence.

\begin{figure} \centering
\includegraphics[width=0.7\textwidth]{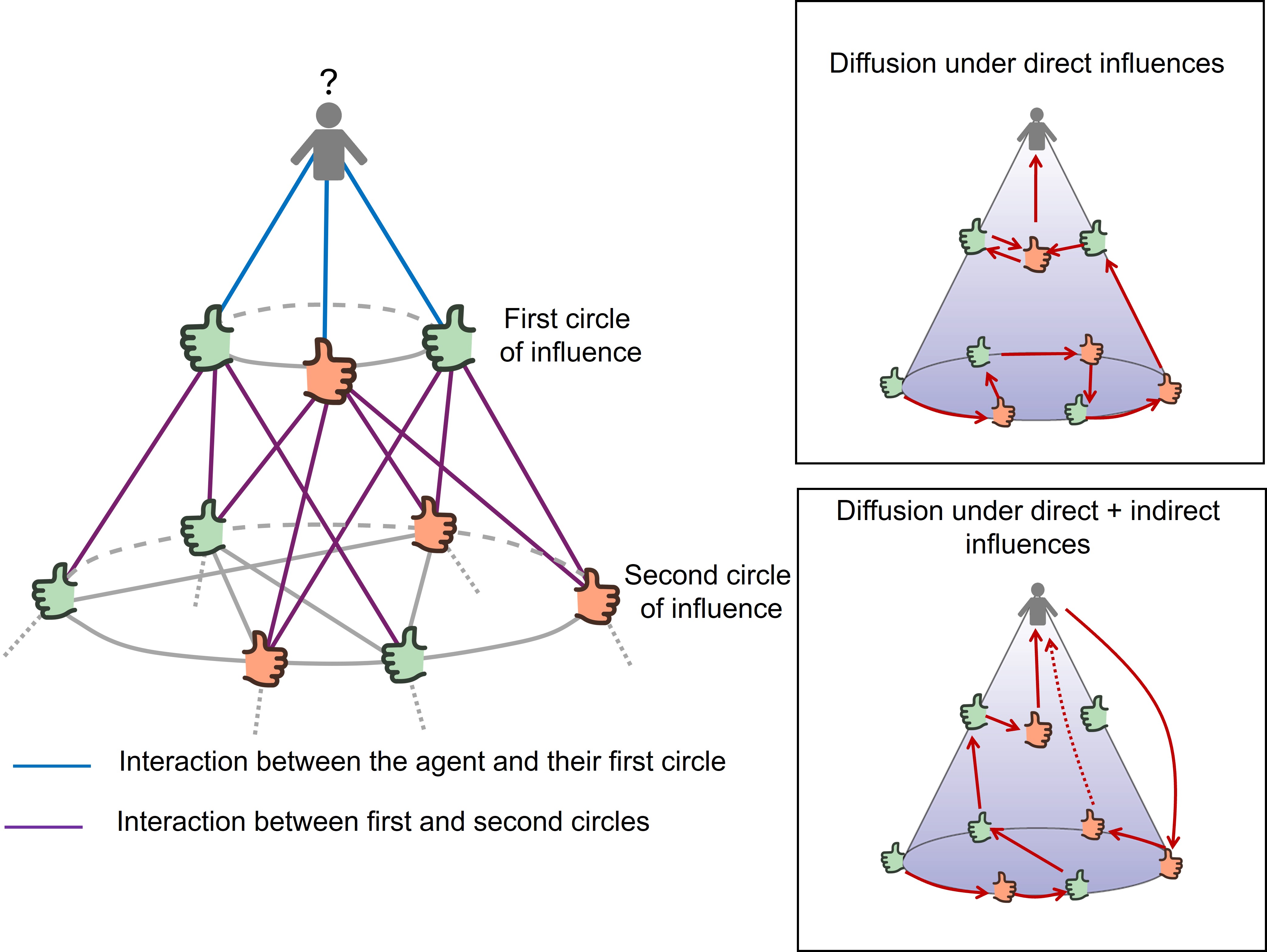}
\caption{Representation of the cone of influences of an individual.}
\label{circle of influences}
\end{figure}

In this work, we analyze innovation diffusion by considering a situation in which an agent, who is influenced by
their interpersonal ties in a social network, is also influenced by
any other individual in the network with a strength that decays with
the social distance separating them. Here, we consider the social distance
to be exactly the number of ties that separate two individuals in
their network of interpersonal channels. That is, the individuals
connected to a given agent form their first circle of influences (see
panel A in Fig. \ref{circle of influences}). Those individuals separated
by two ties form the second circle of influences, and so forth. Then,
the traditional view of the process of diffusion in which an innovation
is communicated through several iterations between agents directly
interconnected by interpersonal ties, for instance, from A to B and then from B to C and from C to D (see panel B in Fig. \ref{circle of influences}), is confronted with the model in which the diffusion process occurs via direct plus indirect influences, where the last take place
via the second, third, etc. circles of influences (see panel C in
Fig. \ref{circle of influences}), such as A receives direct influence from B, but also indirect influences from C and D. 
The framework for our research is a study of the adoption of a drug between physicians in a hospital \cite{menzel-1955}: based on this work, we designed and conducted a series of experiments to empirically measure how long-distance connections affect the adoption of innovations in a social network. Using the mathematical framework developed in \cite{estrada-2012,estrada-2013,estrada-2017,estrada-2018}, we can measure the strength of such non-direct connections compared to the effect of direct friends in such adoption process. Therefore, we are following Rogers' suggestion that \textquotedblleft Alternative research approaches to post hoc data gathering about how an innovation has diffused should be explored. \textquotedblright \cite{rogers-2003}. While some recent experiments have been done in this field \cite{muchnik-2013,aral-2014,chami-2017,aral-2011A,bapna-2015}, the effect of non-direct relationships has only been studied through data collection of studies not specifically designed to such purpose \cite{bartal-2019,festinger-1952}. Therefore, our work fills the gap of experimental research explicitly designed to address the issue of non-direct connections on innovation diffusion.

\section{Experiment and model}

\subsection*{Experimental Setup}

As stated above, to investigate the existence and strength of influences--direct influences only or direct plus indirect ones--on the adoption of an innovation, we designed the following experiment. A group of participants is placed in all nodes of a network. At each round of the experiment, every participant has to choose between two colors, one which is assigned to the majority of participants and another one assigned only to a small number of them (but the respective fractions are not known to the participants). The participants receive a monetary reward (see Methods section \ref{experimentmethods} for details) if they reach a global consensus in one of the two colors. The incentive is inversely proportional to the number of rounds they
need to reach this steady state. In this scenario, in the initial round, the color of the majority represents the ``tradition'',
while the color of the minority represents an ``innovation''. To stimulate the adoption of the innovation, participants receive a greater incentive if a consensus is reached on the initial minority color. Although the proportion of vertices with each of these colors evolves with time, we will refer to ``majority'' and ``minority'' throughout the experiment. In each session of the experiment we randomly assigned some pairs of participants to role as friends. Pairs of friends form the edges which are created by imitating one network previously studied for the diffusion of an innovation in the real-world \cite{menzel-1955}. 

\begin{figure*} \centering
\includegraphics[width=\textwidth]{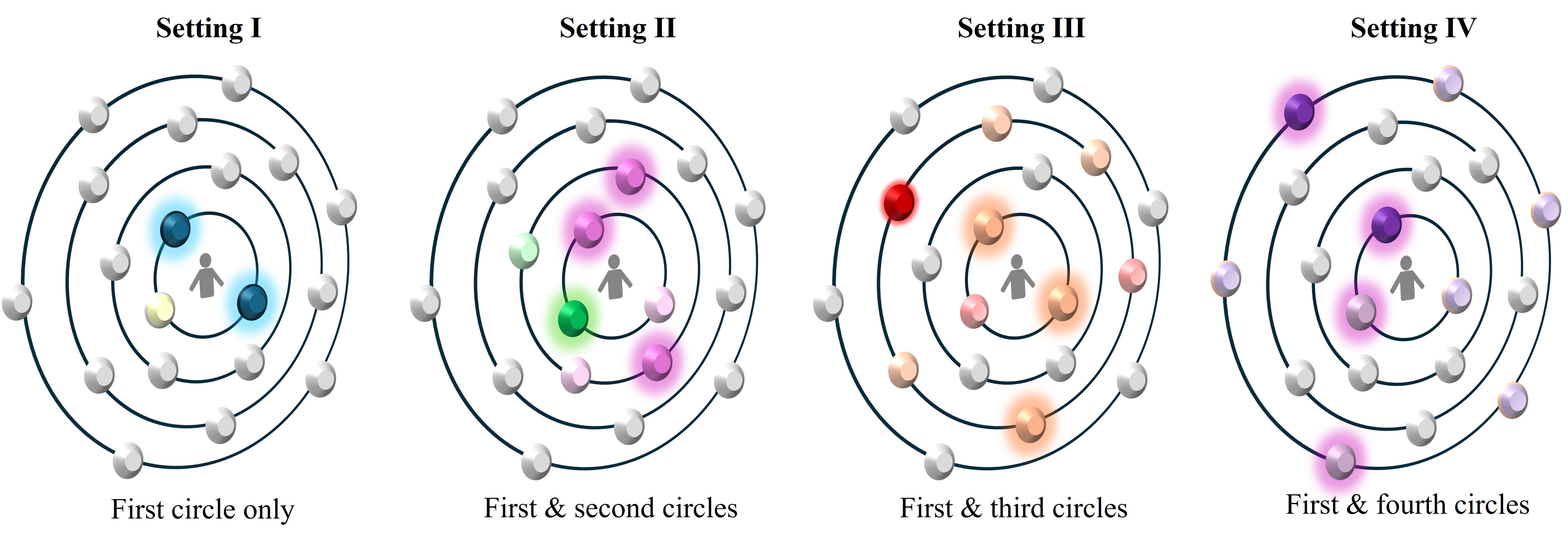}
\caption{Schematic illustration of the experimental setup used in this work.
The first setting consists in showing the participants information
about the color chosen by two other subjects directly connected to
them in the network. In the $k$-th settings ($k=2,3,4$ for settings
II, III, IV, respectively) the subjects observe the colors of two
participants connected to them and two others in their $k$-th influence
circle. The colors in every setting change to about bias in the color
selection: setting I (blue and yellow); setting II (magenta and green);
setting III (orange and red); setting IV (purple and lilac). The first
color in every pair is always the one of the \textquotedblleft minority \textquotedblright.
In the figure, the vertices observed by the subject are \textquotedblleft illuminated\textquotedblright ,
while the others in the same circle are shown with pale colors (they
are not observed at all by the subjects).}
\label{Experiment}
\end{figure*}

Let us now discuss how the initial choice of colors is assigned to the participants. 
Assuming that only 13\% of the experimental subjects were early adopters, we initialized all sessions with 27 participants having the majority
color and only 4 with the minority one. Once the participants have been assigned a color, the experiment takes place in four different settings. In each one of them, every subject sees a picture of
the network, presented as an ego network centered at themselves, with
vertices at longer distances being smaller than the ones which are
nearer to them. The picture they observe is similar to the
one in Fig. \ref{Experiment} where only limited information about
the colors that every other vertex has is provided to them. The exact screenshots of the experiment can be found in the Supplementary Material. A round finishes when all participants have made their choice. At
the end of every round,
participants see again a similar picture with updated information about the colors
of the corresponding neighbors. 

In setting I, every subject had information about the colors of two
of the agent's nearest neighbors. In setting II, such information
consisted of the colors of two of the agent's nearest neighbors and
two of the participants who are at distance two from the target. Similarly,
for settings III and IV, the information provided was about two nearest
neighbors and two in layer three or four, respectively (cf.\ Fig. \ref{Experiment}). Every experimental session consists of a sequence of the four settings, each one comprising in turn of 13-15 rounds. A round is defined as the step in which every participant makes a decision, either keeping or changing the color currently assigned to that agent. Although participants knew that settings would end after at most 15 rounds, they did not know the exact maximum number of rounds used in each
setting. In each session, the order of the settings was randomized to avoid order effects. Further details of the experimental design can be found under Methods, section \ref{experimentmethods}, and the Supplementary Material.

\subsection*{Theoretical Model}

In order to understand and analyze our experiments, we will compare them with the following analytical model \cite{estrada-2012,estrada-2013,estrada-2017,estrada-2018}. 
Let us assume that every agent $i$ has a propensity
$u_{i}(0)=u_{i}^{0}$ to adopt an innovation at an initial
time $t=0$. Then, the adoption of this innovation in a network is
a consensus process in which the change of the state of subject $i$
at a given time, $\dot{u_{i}}(t)$, is determined by

\begin{equation}
\dot{u_{i}}(t)=\gamma_{NN}\sum_{NN}\left[u_{j}(t)-u_{i}(t)\right],
\end{equation}
where $\gamma_{NN}$ represents the ``strength'' of the nearest
neighbors (NN) interactions, i.e., those vertices in the network directly
connected among them. If we represent the states of every individual
at a given time in the vector $u(t)$, we can write

\begin{equation}
\dot{u}(t)=-\gamma_{NN}\mathcal{L}_{NN}u(t);\:u(0)=u^{0},
\end{equation}
where $\mathcal{L}_{NN}$ is the Laplacian matrix of the network operating
over the pairs of NN individuals. This is understood as an operator
on a Hilbert space on the set of vertices of the network acting on
a function $f$ defined in the same space and evaluated on the vertex
$v$ as: $(\mathcal{L}_{NN}f)(v):=\sum_{NN}\left[f(w)-f(v)\right]$,
where the sum is over all the NN of $v$.

The solution of this equation is: $u(t)=e^{-t\gamma_{NN}\mathcal{L}_{NN}}u^{0}$,
and the steady state is the one in which every vertex has a state
equal to the average of the initial condition. This equation represents
the diffusion of the adoption of the innovation across the network,
assuming that the process is continuous in time as well as in which
the state of the individuals may take a continuous range of values.
However, in an experimental setup as the one described before the
time is discrete as it is determined by the rounds taken to reach
the consensus and the results are binary, i.e., an agent either adopt
or does not adopt the innovation. Therefore, here we discretize time
as follows. For any given time $T$ and a number of rounds $r,$ we
equidistribute $r$ points in the interval $\left[0,T\right]$, such
that the discretized solution is equal to the continuous solution
at those times. We also proceed to discretize the output of the model
by introducing a threshold parameter: $\triangle\cdot u_{i}(t_{c})$,
where $\triangle\in\left[0,1\right]$ and $u_{i}(t_{c})$
is the state of the vertex $i$ when the consensus was reached, which
is equal to the average of the entries of $u^{0}$. This means that
when an agent has a propensity to adopt the innovation larger than
this threshold it is assumed that the agent adopts the innovation.
Otherwise, it is assumed that it has not adopted the innovation. 

To account for the influence of the individuals in the second circle
of influence of the agent $i$ we can define the Laplacian operator
$(\mathcal{L}_{NNN}f)(i) := \sum_{NNN}\left[f(j)-f(i)\right]$,
where now the sum is carried out over all next nearest neighbors (NNN)
of $i$, i.e., those separated by two edges in the network. Similarly,
we can extend this definition to the third, fourth and so for NN of
a given agent, such that we can write the innovation diffusion model
as:

\begin{align}
\dot{u}(t)=-\gamma_{NN}\mathcal{L}_{NN}u(t)-\gamma_{NNN}\mathcal{L}_{NNN}u(t)-\cdots -\gamma_{D}\mathcal{L}_{D}u(t); \qquad u(0) = u^0,
\end{align}
where $\gamma_{NNN}$ is the ``strength'' of the interactions between
next nearest neighbors and $D$ designates the diameter of the network,
i.e., the longest shortest path between any pair of vertices. The
intuition dictates that the strength of the interaction decays with
the separation between the pairs of agents in the network, i.e., $\gamma_{NN}>\gamma_{NNN}>\cdots>\gamma_{D}$.
In the experiments designed in this work the diameter of the network
is five, i.e., $D=5,$ and we can use the following notation accordingly:
$c_{1}=\gamma_{NN};c_{2}=\gamma_{NNN};\ldots$.
Similarly, we designate $\mathcal{L}_{1}=\mathcal{L}_{NN}$; $\mathcal{L}_{2}=\mathcal{L}_{NNN}$,
etc., where, as defined before, $(\mathcal{L}_{d}f)(v) := \sum_{d(v.w)=d}\left[f(w)-f(v)\right]$.
Let us fit $\gamma_{NN}=c_{1}=1$, such that we can write:

\begin{equation}
\dot{u}(t)=-(\mathcal{L}_{1}+\sum_{d=2}^D c_{d}\mathcal{L}_{d})u(t);\:u(0)=u^{0}.\label{eq:Model}
\end{equation}

\section{Experimental results}

We recruited 592 participants from the IBSEN subject pool at Universidad Carlos III de Madrid (UC3M) to participate in a series of 21 experimental sessions. The research was approved by the Ethics Committee of UC3M and was carried out with the approved plan. 
The average age was 30.4 years (median 25, mode 22). The gender representation was 63.8\% female, 35.9\% male, and 0.3\% non-binary. The distribution of gender and age through the experimental sessions is shown in the Supplementary Information in Table S1 and Figs. S14 and S15.

To analyze the experimental results from a realistic perspective, we considered that a subject
who has adopted the ``minority'' color becomes an adopter of the innovation
from that round on.  This definition is intended to take into account the differences in time between
the experimental settings and the real adoption of an innovation.
While the first takes minutes, the second can take years, and once
a subject has adopted an innovation in the real world, it will take
long times until they can abandon it, in case they ever do so.
In 14 of the 21 experimental sessions, the participants reached
consensus in setting I, and for settings II-IV, the global consensus
was reached in 16 of the 21 sessions (see Fig. S9 in the Supplementary Information). This means
that in some of the 21 experiments there was at least one individual
(stubborn) who did not join the consensus of the group for the duration of the setting. In total there were 11, 6, 10 and 8 stubborn individuals
in settings I, II, III and IV, respectively, which clearly points
out to the lack of any bias in the number of such individuals in relation
to the type of social interactions considered in the experiments.

\begin{figure} \centering
\begin{centering}
\includegraphics[width=0.7\textwidth]{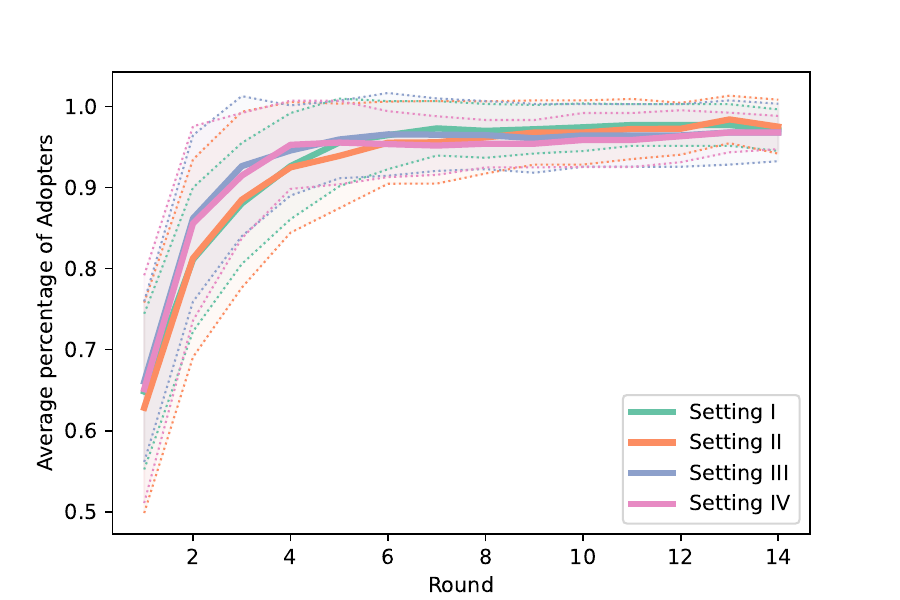}
\par\end{centering}
\caption{Average percentage of adopters (including bots) as a function of the experimental rounds, per treatment. Shaded areas represent standard deviations.}
\label{adoption rate}
\end{figure}

In Fig. \ref{adoption rate} we illustrate the cumulative distributions of the proportion
of adopters of the innovation versus round for each of the four settings
considered here and averaged over the 21 experimental sessions.In sessions where most participants reached consensus in round x but some were stubborn, we adjust the Fig. \ref{adoption rate} plot to show global consensus at round x+2 for aesthetic purposes. In general, the percentages of adopters in the second
round are approximately the same for the four settings (for round
I all the percentages are exactly the same as we initialize all the
experiments with this percentage of adopters). However, for rounds
3-5 these percentages show the largest differences between the four
settings. In round 3, the percentage of adopters in setting I is about
84.5\%, while for settings II and III it grows to 88.8\%, and for
setting IV it is 85.4\%. In round 4 these percentages are: 89.2\%,
92.9\%, 93.4\% and 91.1\%, respectively, and for round 5 they are:
92.9\%, 96.2\%, 96.2\% and 94.2\%. Although these values are average
percentages that may be hiding the specifics of each session,
(see further analysis), they clearly indicate an acceleration in the
number of adopters in settings II-IV relative to setting I, particularly
for settings II and III. That is, these results seem to point
to the fact that the second and third neighbors of an agent significantly
influence their decision in choosing an innovation. Such influence
seems to drop for the fourth circle of influences.

Let us now discuss the fit of 
the experimental results to our model. To that end, we proceed by considering the individual experiments. For each setting, we fit the results of each of the experimental
sessions to find the parameters $c_{2},$ $c_{3}$ and $c_{4}$ as
well as to find the values of $\triangle$ (the value of $u(t)$
that triggers the adoption of the innovation) and $T$ (the time equivalent
to the number of rounds in the experiment) that best fit the data
as detailed in Methods section \ref{FitMethod}. 

\begin{figure}[t] \centering
\begin{centering}
\includegraphics[width=0.8\textwidth]{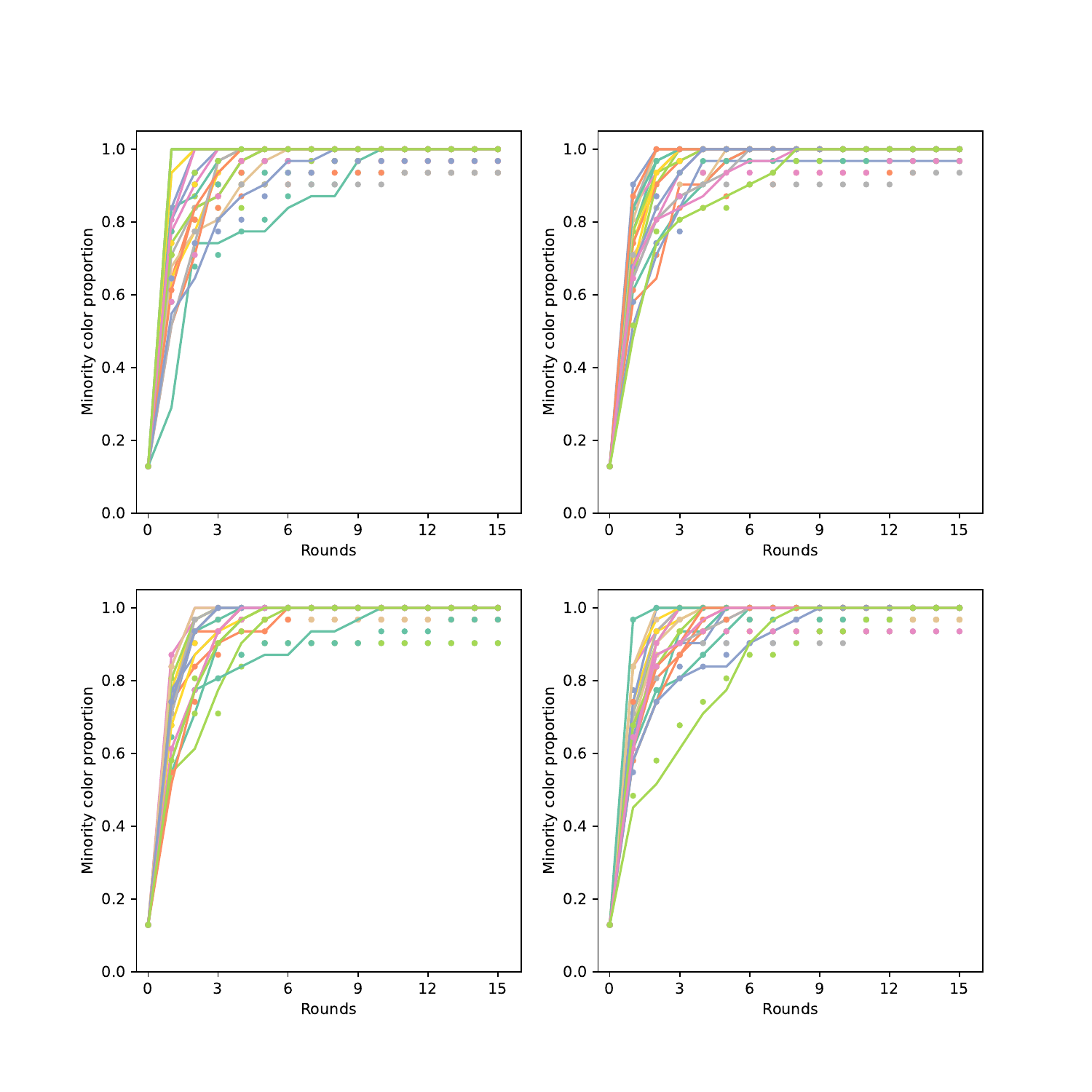}
\par\end{centering}
\caption{Proportion of adopters (dots) and their theoretically expected behavior (solid lines) as a function of the rounds for all the experimental sessions (in colors) and the different settings: setting I in upper left subplot, setting II in upper right subplot, setting III in lower left subplot and setting IV in lower right subplot.}

\label{Fittings}
\end{figure}

In Fig. \ref{Fittings} we illustrate the results of the fitting procedure
for the four settings in the 21 experiments. The experimental data
is visualized as points of colors representing each experimental
session. The best fits obtained by the procedure described previously
are illustrated as curves of the same colors as those of the data points.
As can be seen in the plots, the fittings are much better for the
initial times of the time evolution of the adoption procedure than
for the final ones. The reason is that, as mentioned before, in several
experimental sessions, there were stubborn participants who never
joined the consensus state followed by the large majority of subjects.
On the contrary, the diffusion model assumes that every participant is predisposed to reach the consensus state. In any event, we have analyzed our experimental data by removing outliers that are basically coincident with the presence of stubborn subjects, and the results are approximately the same (see the Methods section \ref{stubborns}). As there are theoretical
models that take into account the presence of stubborn participants, we maintain
the general idea of using a diffusion model as our main goal here
is to investigate the role of indirect peers pressure in the adoption
of innovations. Further studies can be designed to design models
in which stubborn participants are explicitly considered.

\begin{figure} \centering
\centering
\includegraphics[width=\textwidth]{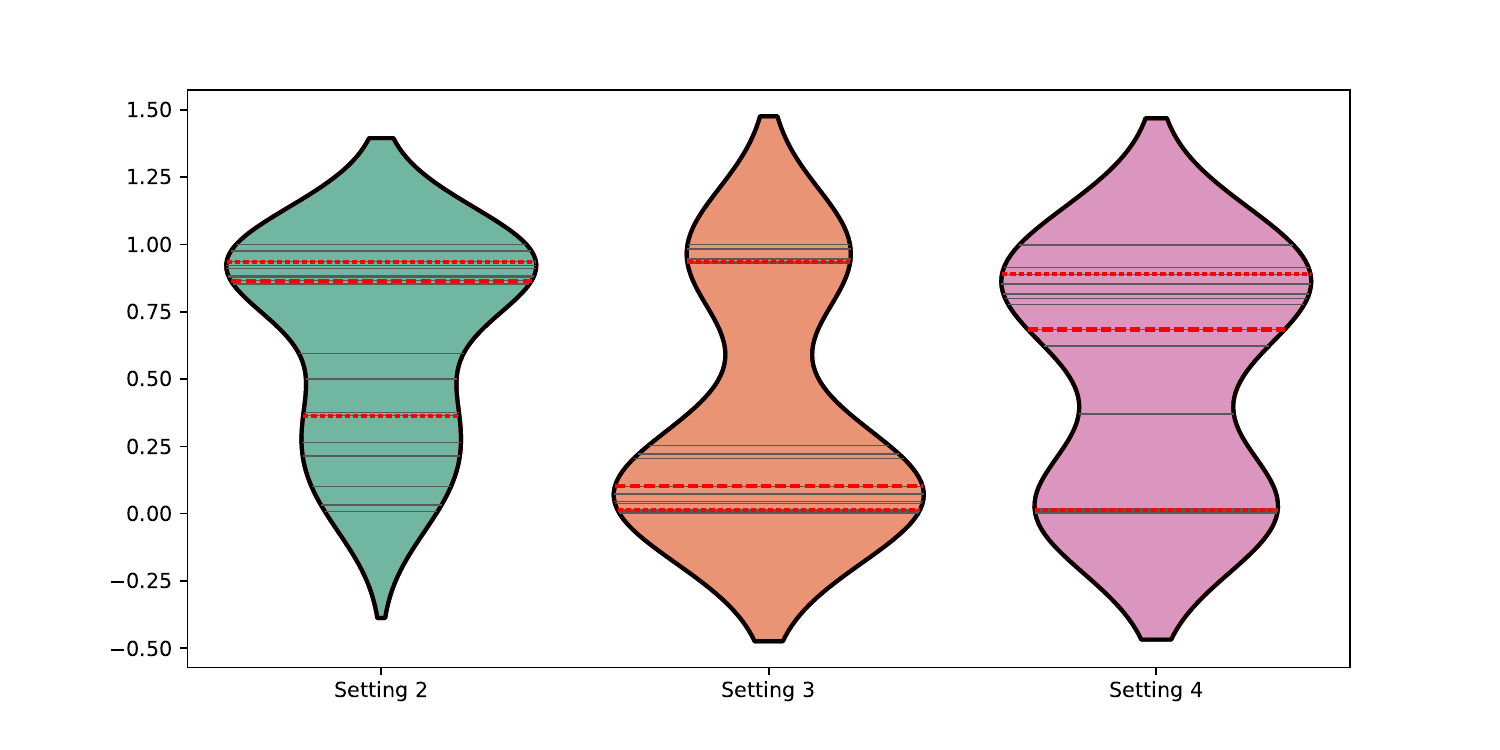}
\caption{Violin plot of the distribution of fitted parameters for each of the Settings that include indirect peer-pressure. Black lines are the individual value of the fitted parameter for each of the sessions, while the red lines are the interquartile ranges of each distribution.}
\label{Parameters}
\end{figure}

From the perspective of accounting for the direct and indirect influences
of peers on the adoption of an innovation, the parameters $c_{d}$
are the most relevant. In Fig. \ref{Parameters} we illustrate
the distributions of the parameters $c_{2}$ (setting II), $c_{3}$
(setting III), and $c_{4}$ (setting IV), obtained from the best fittings
of the experimental data to the models of direct plus indirect influences
on the network. The values of the mean and standard deviations of
these coefficients are as follows: $c_{2}=0.651\pm0.354$; $c_{3}=0.373\pm0.427$;
$c_{4}=0.513\pm0.420$. We recall that the strength of direct
influences is $c_{1}=1$.

We then check whether the differences
between the means of these coefficients are significant according
to their $P$-values, i.e., the probability of obtaining the observed
difference between the samples if the null hypotheses were true. The
null hypothesis states that the difference between the averages is 0, that is, there is no difference. We obtained: $p(c_{2},c_{3})=0.0269$, $p(c_{2},c_{4})=0.256$,
$p(c_{3},c_{4})=0.290$. Therefore, the only significant
difference, i.e., $p<0.05$, is between the coefficients that represent
the influences of the second and third circles, but not between the second or third with the fourth, where there is no empirical evidence to reject the null hypothesis.

This lack of significant difference
between the means of $c_{4}$ and the other two coefficients could be
due to several experimental factors that cannot be explained with
the information that we have obtained from them. Consequently, we
eliminate the results concerning the influence of the fourth circle
of influence and focus on the fact that our results indicate that
there is a relatively large influence of the second NN on the adoption
of an innovation by an agent, which is on average 65\% as strong as
the direct influence of peers, and a relatively small, but not negligible,
influence of the third NN, which is on average 37\% as strong as the
direct interaction. We can then write an approximate model that describes
the results of our experiments as follows:

\begin{align}
\dot{u}(t)\approx&-(\mathcal{L}_{1}+(0.651\pm0.354)\mathcal{L}_{2}+(0.373\pm0.427)\mathcal{L}_{3})u(t); \qquad u(0) = u^0. \label{eq:empirical}
\end{align}

The empirical model (\ref{eq:empirical}) reflects the fact that the
strength of the influences decays with the increase in the social
distance (measured here as the shortest path) between the subjects.
This model can be approximated very well by considering that the coefficients
$c_{d}$ are indeed a linear function of the distance, such that we
can write:

\begin{equation}
\dot{u}(t)\approx-\sum_{d=1}^{D}\left[(\dfrac{4-d}{3})\mathcal{L}_{d}\right]u(t);\qquad \:u(0)=u^{0}.
\end{equation}

Using this approximation we can say that the strength of the interactions
between a subject and its second circle of influences is about 2/3
of that with their closest neighbors, and those in the third circle
have an influence which is about 1/3 of the ones between NN.
Whether
this is a general expression for other cases of diffusive adoption
of innovations is something which should be taken with prudence and
analyzed in individual cases.
Linear decay models have previously been used to consider social effects,
such as rumor transmission in a network, where an exponentially truncated
linear decay function is used to characterize the decay such that if
the acceptance time is small, the decay function is dominated by a
linear decay function \cite{wang-2019} (see also \cite{kempe-2003,domingos-2001}). Other kinds of decay are also studied in the Supplementary Information.

To gather further evidence on the role of non-direct influences, we now focus on the individual decisions. We want to unveil which pieces of information are people using to make their decisions (whether to choose the innovation color or not). In order to do so, we study the problem as a classification problem, where our aim is to predict the color a person chose as function of the information available: whether people had the innovation as their initial color, whether the innovation color is the majority color seen, the percentage of their first neighbors with the innovation color, and the percentage of $n$-distance neighbors with the innovation color. Then, the chosen color is the dependent variable and the four pieces of information are the features or independent variables. We use Random Forests as classification technique and the feature importance analysis (see the section of Methods \ref{RF} for details).

\begin{figure} \centering
    \centering
    \includegraphics[width=0.7\linewidth]{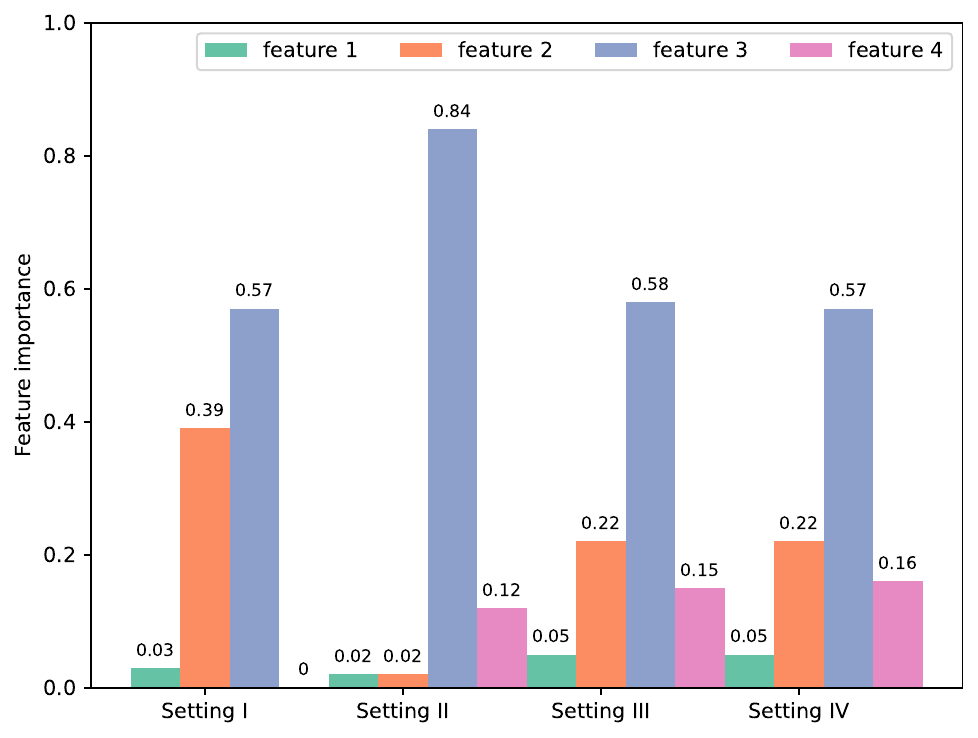}
    \caption{Feature importance analysis per experimental setting. The features are: (1) innovation as initial color, (2) innovation as majority color seen, (3) percentage of first neighbors with the innovation color, and (4) percentage of n-distance neighbors with the innovation color.}
    \label{fig:featureimportance}
\end{figure}

Our first result is that the color that a person chooses in each round can be predicted with high accuracy ($84\%$). Subsequently, when we study the feature importance of this classification problem, i.e., the contribution of each piece of information to predict people's decisions, we see that the importance of the four pieces of information is different depending on the experimental setting (see Fig. \ref{fig:featureimportance}).

As can be seen in the plot, the initial color assigned to the participants (feature 1) is irrelevant to people's decisions. 
In the first setting,  participants have no information of their neighbors at distances bigger than one, and hence feature is being unimportant in this setting. The most important feature is whether the direct neighbors have acquired the innovation or not. It is twice as important as the majority opinion (feature 2). In settings II to IV, the information of $n$-distance neighbors is relevant for the decisions, and, notably, this information becomes very important. 
In setting II, with direct neighbors and neighbors at distance two, the most important variable is still the direct neighbors information; four times more important than the 2-distance information. In this setting, the information of the majority (feature 3) is irrelevant.
In settings III and IV, there is a decrease in the importance of the first neighbors information in favor of the importance of the majority and the $n$-distance neighbors information. This suggests that people are taking the whole picture into account when making their decisions. In general, the analysis suggests that the $n$-distance information influences the adoption of innovations, and this is more relevant as the information present is from higher distances.

\section{Discussion}
\begin{figure}[t] \centering
    \centering
        \includegraphics[width=0.85\textwidth]{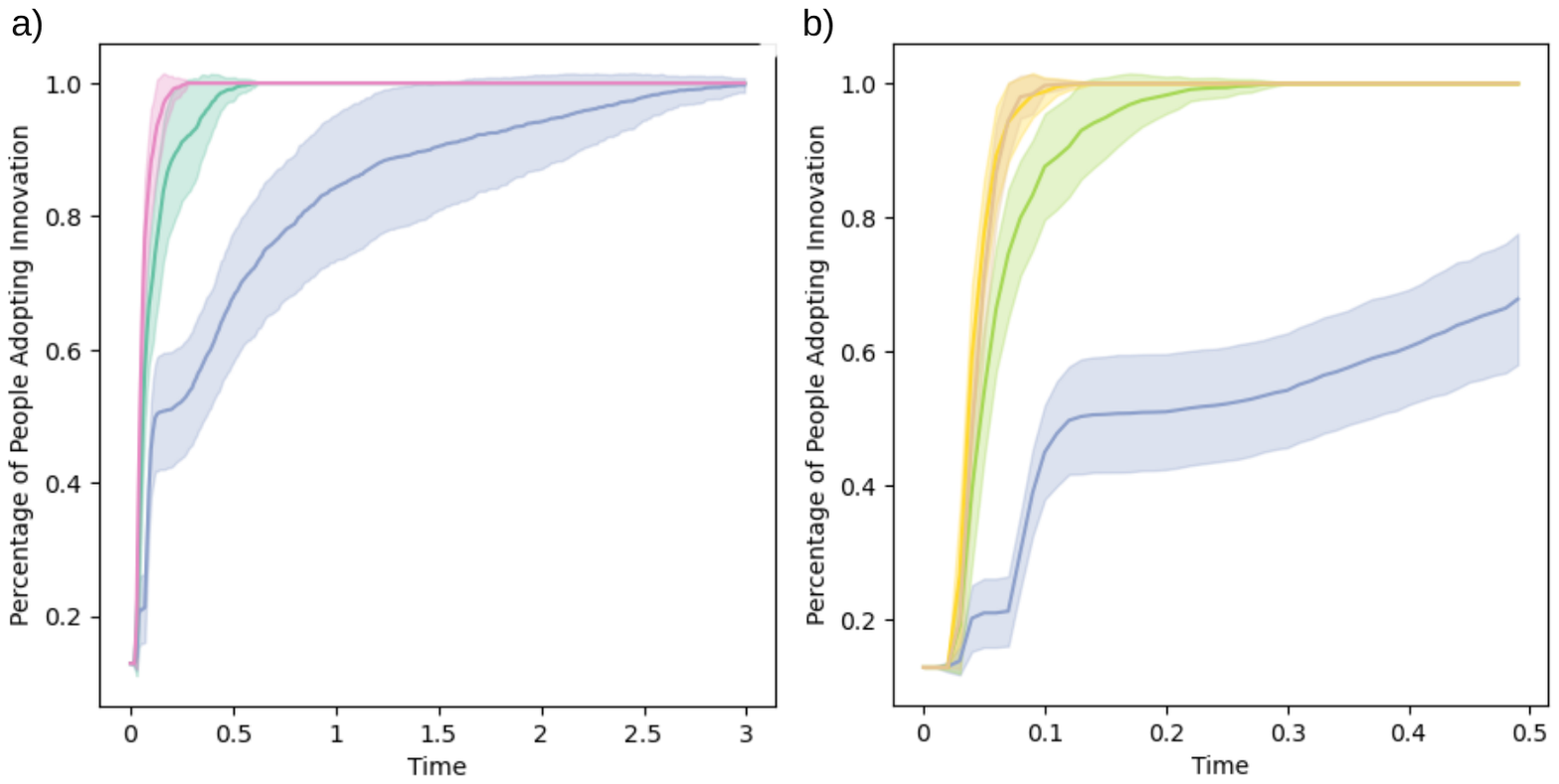}
        \caption{Comparison between different laws for indirect peer pressure, averaging over 1000 initial conditions. Blue line represents the only direct pressure diffusion. Dark green line corresponds to considering also indirect pressure from next nearest neighbors, while pink line considers pressure from all neighbors at distance less or equal than 3. Light green line corresponds to the law $c_d = \frac{4-d}{3}$ commented in section 2.A. Yellow and brown lines represent two laws that accelerate the spreading of the innovation, using the laws $c_d = \frac{5-d}{4}$ and $c_d = \frac{2}{1+d}$, respectively.}
        \label{Accelerating diffusion}
\end{figure}

In this paper, we have provided solid evidence pointing out to the fact that indirect influences play a fundamental role in innovation diffusion, a key process in a globalized technological society such as the current one. The results of an experiment specifically designed to probe into this question demonstrate that, as summarized in Fig. \ref{Accelerating diffusion}, the adoption of the innovation by about 60\% of participants in our experiments may take around five times less steps if we
allow them to see the influence of those socially close but not
connected to them. The situation is even more dramatic if we consider the times at which 80\% of the experimental subjects adopted the innovation. In this case, the reduction of time is more than 10-fold
under the indirect influence of peers: In practical terms, this means that an innovation which would
take around a year to be adopted under direct influences only, would be adopted in about one month
under the joint effect of direct and indirect influences. We note also that the diffusion of innovation goes faster in the first stages of the diffusion process if information on long-range distance is present; see setting III and IV curves in Fig.~\ref{adoption rate}. Further, independent evidence that information is indeed the mechanism behind the acceleration of innovation diffusion comes from a feature analysis that reveals the way participants weigh their knowledge of the social context. All in all, the experimental evidence sends a clear message with practical implications: innovation diffusion can only be properly understood if the influence of people at different levels of social distance is taken into account. 

As an additional illustration of how such indirect influences can change the rate of diffusion of an innovation, we do the following theoretical experiment. By considering the same network studied here
experimentally, we tune the indirect influences without changing the direct interactions between the
agents. We tune these influences simply by changing the coefficient $c_d$ which determines the weight
that the not-direct influences have on a given agent. In Fig. \ref{Accelerating diffusion} we illustrate the results where we also
plot the experimental results obtained here for no indirect influences as well as for direct+indirect
ones, in which the coefficient fitted to the experimental data is $c_d = \frac{4-d}{3}$. When we increase
such indirect influences to $c_d = \frac{5-d}{4}$ or to $c_d = \frac{2}{1+d}$, the results are obvious: a significant
increase of the diffusive dynamics in which the times to adopt the innovation are significantly
reduced.

One interesting question arising from our experimental results is the lack of differences between the coefficients for the fourth neighbor influence and the coefficient for the second and third neighbors. While, as already mentioned, this may be an effect of sample size or, perhaps, of the network size, it may also be the case that the weight we give to the influence of socially distant contacts saturates, meaning that beyond the first two or three layers of contacts we take in the input of further ones in the same manner. This might arise as a consequence of limited cognitive capabilities: in a general situation in the population at large, we will have many more contacts as social distance increases, and we are thus led to consider them in a less specific manner. Remarkably, these results coincide with those reported by Christakis and coworkers who observed that the risks of spreading obesity \cite{christakis-2007}, smoking behavior \cite{christakis-2008}, happiness \cite{fowler-2008A}, and alcohol consumption behavior \cite{rosenquist-2010} are influenced by individuals up to three degrees of separation between each other. They observed that by the fourth degree of separation there was no excess relationship between individuals in the large social network analyzed over a period of 32 years. Further experiments in larger networks would help clarify this point, although it must be taken into account that that would require a large sample of subjects that would play simultaneously with the logistic challenge that implies \cite{pereda-2019}.

In terms of real-world impact, our results indicate that acting on indirect influences
could change very significantly the adoption rates of innovations. Mass media has been frequently
identified as a principal actor of indirect influences. By this means, for instance, teenagers in one
country can observe the attitudes and behaviors of others in a different one, copying them for good or for bad. Therefore, mass media can act as a modulator of indirect influences, which may change
significantly the dynamics of innovation adoption. While this is a source of influence which is not really amenable to use as an intervention, other approaches may lead to specific actions in order to increase, e.g., the adoption of socially desirable behaviors. Akin to behavioral interventions in which people are informed of the expectations of others on their behavior \cite{bicchieri-2016}, we could think of campaigns in which subjects of interest receive information on what other, socially distant, people do in the relevant context. Our experimental results indicate that giving only a limited amount of information about second- or third-order contacts could already lead to highly increased rates of behavior adoption. 

In closing, the results found in this work clearly point out to the fact that when adopting an
innovation we are not only influenced by peers directly connected to us in our social networks, but
that we are also significantly influenced by people socially close but not directly connected to us in
any of our social networks. This work paves the way to the development of further experimental and theoretical setting which will allow us a better understanding of the innovation diffusion dynamics.

\section{Methods}
\subsection{Experimental methods} \label{experimentmethods}

The experiment consisted of four treatments, which we refer to as ``settings'' to highlight our interest in informational settings, to study the influence of peer pressure on consensus. Here we summarize a few additional details that are needed to complete the definition of the experimental setup. 
The colors in each treatment were different to avoid learning biases. The eight colors were: setting I (blue and yellow); setting II (magenta and green); setting III (orange and red); setting IV (purple and lilac). The first color of each pair being the majority. 
In case one or more subjects
did not make a choice in a given round, we declared them ``inactive''.
Then, to avoid any change in the structure of the underlying network,
we replace that player(s) with a ``bot'', which is programmed to
have 50\% probability of choosing color at random and 50\% of following
the majority of the color they would see. Nonetheless, bots were clearly
marked as inactive subjects and were not shown to active participants
if possible. The original network from the study \cite{menzel-1955} was slightly modified so that every node had at least two nodes at distance one and two nodes at distance four. This was done by removing the edge from node 10 to node 30 and by adding an edge between nodes 2 and 17.

Participants received points that, at the end, were converted into money. Each participant received 1 point per active round (if they made a decision before the timeout occurred). Then, if consensus was achieved, all of them received 5 points per each round left until round 15 (maximum possible number of rounds) if they were active at least in one of the two last played rounds of that setting.

These experiments were programmed using the Python package oTree \cite{oTree} version 5.10.3, using Cytoscape.js \cite{cytoscape} version 3.1.0 for graph visualization. The code is available in \cite{githubrepo}. Data results are available in \cite{Zenodorepo}. Snapshots of the webpages presented to the participants are shown in the Supplementary Information.

\subsection{Fitting method} \label{FitMethod}
For a given experiment and a specific setting, we obtain the vector
$u_{exp}$ with the percentages of adopters in each round, as well
as the initial condition vector $u_{0}$. Then, for each triplet of parameters $(c_d,\triangle,T)$, we produce a prediction vector $u_{pred}$ based on our model using the following method.
From the interval $[0,1]$, discretized in intervals of size $0.01$, we choose $c_d$ and calculate the solution of the dynamical system, obtaining the solution $u(t)=\left[\exp(-t(\mathcal{L}_{1}+c_{d}\mathcal{L}_{d}))\right]u_{0}$. Now, we fix a value for $T$ taken between $10$ and $1000$ in steps of size $10$, and we identify each round $i$ of the experiment with time $t_i \in [0,T]$ such that these times are equally distributed throughout the interval. Finally, we use one of the threshold values $\triangle \in \left\{ 0.4,0.5,0.6,0.7,0.8,0.9,0.99\right\}$ and calculate each entry of the vector $u_{pred}$ by summing all the entries of $u(t_i)$ above the threshold ($u_{pred} (i) = \emph{sum} (u(t_i) > \triangle)$.
Hence, for each combination of $(c_d,\triangle,T)$, we obtained the prediction $u_{pred}$, which we can compare to the real proportion of adopters during the experiments ($u_{exp})$ by calculating the mean square
error (MSE) and the Pearson correlation coefficient between the two vectors. The best fit is the combination of parameters that minimize MSE.

\subsection{Random Forests and feature importance analysis} \label{RF}
Random Forests (RF)~\cite{Breiman2001} are one of the most effective algorithms, excelling in predictive performance in various application domains, while demonstrating robustness against overfitting and internal correlations among explanatory variables. Random Forests employ decision trees with a unique ensemble technique called bootstrap aggregation (bagging). Unlike traditional decision trees, RF combines the results of multiple weak learners using bagging, which aggregates results through averaging in regression tasks and a voting system in classification tasks. One notable feature of RF is its utilization of "Out-Of-Bag" (OOB) data, which comprises approximately one-third of the original dataset that is not used in constructing each tree. OOB data serve as a test data set to estimate misclassification error and can also be used to analyze the relative importance of each feature in the classification problem. For every tree in the forest, the $jth$ feature of the OOB sample is randomly permuted, and the resulting increase in OOB error is computed. This increase serves as a measure of the importance of the $jth$ feature for correct classification: the greater the increase in OOB error, the more critical the variable is for achieving accurate classification. This analysis provides valuable insights into the contribution of each variable to the classification process, helping in feature selection and interpretation. To estimate the accuracy of the classifier, we used nested cross-validation (NCV)~\cite{ANDERSSEN200669}. NCV operates by employing cross-validation (CV) within two sequential loops: an inner loop for hyperparameter selection and an outer loop for computing test error. In our experiments, both the inner and the outer loops utilized five-fold CV.

\subsection{Analysis of outliers and stubborn individuals} \label{stubborns}

It is evident that, during the realization of experiments, several uncontrollable
factors may produce outliers which deviate from the statistical behavior
of the majority. In order to detect such individuals, we tested the
results against three different methods for outlier detection: $Z$-score, Tukey method, and mean regression (MR). The first two methods are well documented in the literature \cite{Zscore,Tukey}, so we need to explain the last method. MR consist in measuring how much the mean changes when we remove the current value. High scores corresponds to values that change notoriously the mean and can be considered an outlier.

In Table \ref{outliers} we give the number of experiments
that the corresponding method detected as an outlier. In parentheses,
we give the number of stubborn individuals that were present in such
experiments. Then, in setting II all outliers identified by Tukey
method coincide with those in which there is at least one stubborn individual.
In setting III,  Tukey and MR identify six outlier sessions all
of which have stubborn participants (6 out of 8 in this setting). Finally, the
eight stubborn participants that appear in setting IV are present in the experiments identified by MR as the outlier sessions. Therefore, the statistics
of the results clearly point out the identification of those experiments
in which there are stubborn participants as outliers, which is what
it should be expected by considering that the assumption of the diffusion
model is the whole predisposition of agents to reach the consensus
state.

\begin{table*}
\centering
\begin{tabular}{|c|c|c|c|c|}
\hline 
setting & $Z$-score  & Tukey & MR\tabularnewline
\hline 
\hline 
II & 5 (1); 9 (1) &  2 (1); 5 (1); 9 (1); 16 (2); 20 (1) & 2 (1); 5 (1); 9 (1); 16 (2); 20 (1)\tabularnewline
\hline 
III & 7 (1) & 1 (3); 2 (1); 5 (3); 7 (1) & 1 (3); 2 (1); 5 (3); 7 (1)\tabularnewline
\hline 
IV & none &  none & 3 (1); 6 (1); 9 (0); 16 (2); 20 (3)\tabularnewline
\hline 
\end{tabular}
\caption{Table showing the outlier rounds detected by the different methods. The numbers in parentheses represent the number of stubborn subjects for that session and Setting.}
\label{outliers}
\end{table*}

Once we have identified the statistical outliers in the experiments,
we proceed to recalculate the models after their removal. The new
coefficients are: $c_{2}=0.707\pm0.319$; $c_{3}=0.431\pm0.453$;
$c_{4}=0.503\pm0.411$. We also recalculate the $P$-values for the
three pairs of coefficients and obtain: $P(c_{2},c_{3})=0.0275$,
$P(c_{2},c_{4})=0.0798$, $P(c_{3},c_{4})=0.589$,
which indicates that the means of $c_{3}$ and $c_{4}$ are less
different than before and, although $P(c_{2},c_{4})$ is
significantly smaller than without eliminating outliers, it still
is not significant at 95\% of confidence. The empirical model without
the statistical outliers (excluding those detected by MR) is then:

\begin{align}
\dot{u}(t)\approx-(\mathcal{L}_{1}+(0.707\pm0.319)\mathcal{L}_{2}+ (0.431\pm0.453)\mathcal{L}_{3})u(t); \qquad
u(0) = u^0. \label{eq:empirical-1}
\end{align}
which increases slightly the strength of the influence of the second
and third circle of influence in relation to the case where the outliers
were not removed. All in all, the results indicate that the statistical
outliers do not significantly affect the general findings that long-range
influences play a fundamental role in the diffusion of innovations
across a network of social interactions. Guided by the fact that most
of the outliers are those having stubborn individuals, we conducted a final
calibration of the model by removing all experiments in which there
was at least one of these individuals. The results are very similar
to those obtained when removing the outliers detected by MR and are
not reproduced here.

\subsubsection*{Acknowledgements:}
A. S. acknowledge support from grant PID2022-141802NB-I00 (BASIC) funded by \\ MCIN/AEI/10.13039/501100011033 and by ‘ERDF A way of making Europe’, and from grant MapCDPerNets---Programa Fundamentos de la Fundación BBVA 2022. M.M and E.E. acknowledge support from Project OLGRA (PID2019-107603GB-I00) funded by Spanish Ministry of Science and Innovation as well as by the Maria de Maeztu project CEX2021-001164-M funded by the MCIN/AEI/10.13039/501100011033.

\end{document}


\title{Supplementary Information: \\Indirect social influence and diffusion of innovations: An experimental approach}
\author{Manuel Miranda, María Pereda, Angel Sánchez and Ernesto Estrada}

\maketitle

\section{Experiment description}
Our experiment was divided into 5 sequential phases: Instructions, the first setting (only short-range interactions) and three differently ordered settings (mixing short and long-range interactions).

\begin{figure}[H]
    \centering
    \includegraphics[width = \linewidth]{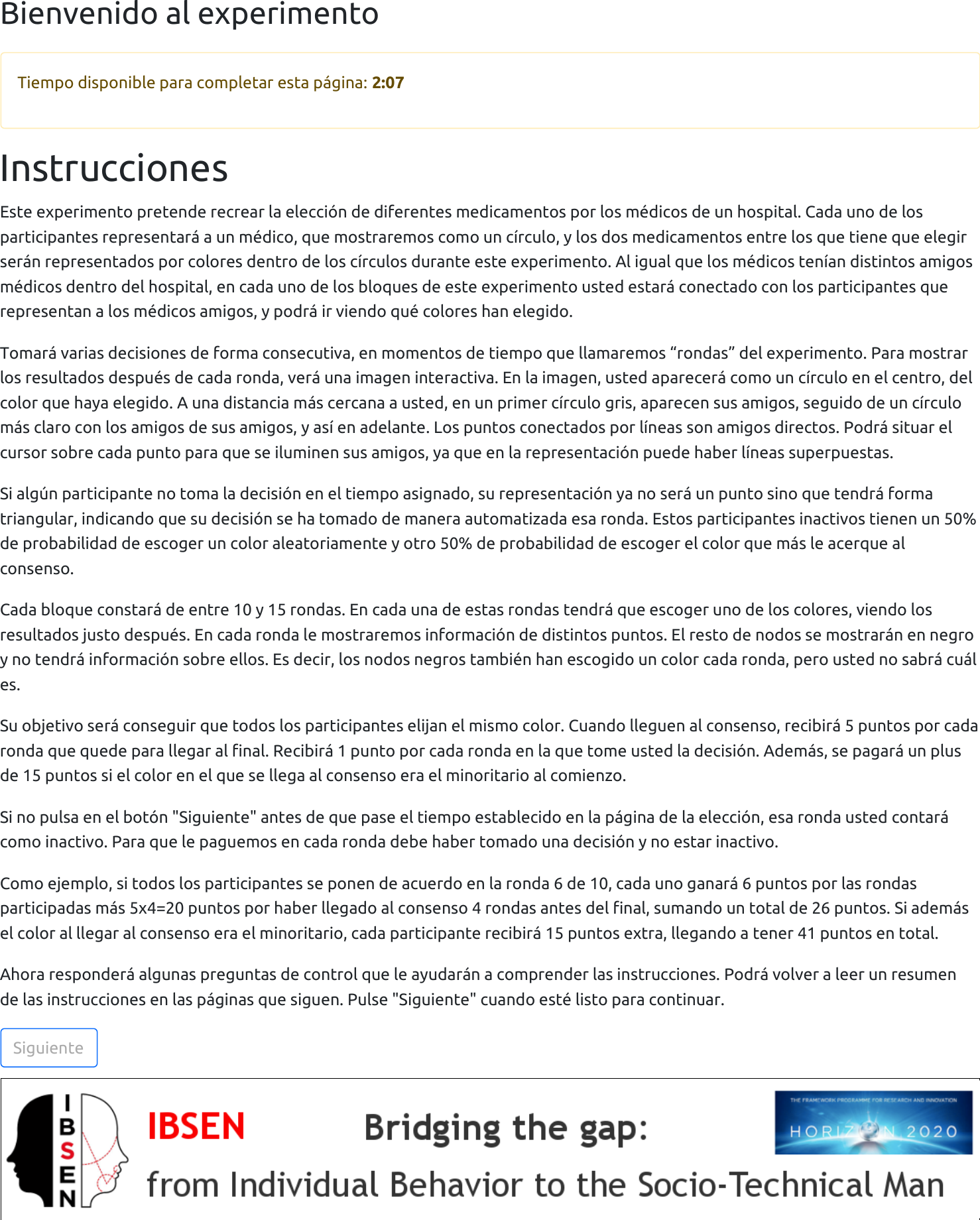}
    \caption{Instructions given at the start of the experiment.}
    \label{fig:Instructions}
\end{figure}

In the Instructions (Fig. \ref{fig:Instructions}), we explained the participants the motivation of the experiment and gave them the rules, objective and payment during the session.
We explained them that our objective was to study how an innovation spreads through a social network, based on the situation studied in \cite{menzel-1955}.In this study,
known as the Columbia University Drug Study, researchers collected
data on the first subscription to the drug ``gammanym'' by 31 physicians
in several communities. The doctors are related via a face-to-face
social network of friendship and discussion, which was created by
asking them to name three doctors whom they considered to be personal
friends and to nominate three doctors with whom they would choose
to discuss medical matters. We explained that, along the experiment, each of the participants would represent one of the doctors from the original study, linked with the same social relations. We made minor modifications to the graph to make sure that every participant has at least two nearest neighbors and that all of them have at least two other participants at distance four. It was needed only to rewire one edge, thus the network has exactly the same density as the original one and about the same topology.

The participants have to choose each round between two different colors, representing two different drugs to prescribe, and one of them will be an innovation which will be a minority at the beginning. To avoid any uncontrollable bias towards the word "innovation", we referred to such color as the minority color at the beginning. The objective of the participants is for all of them to choose the same color as soon as possible. The participant were told that each setting would consist in a number of rounds between $10$ and $15$, although we always chose random numbers between $13$ and $15$. Each participant would win $1$ point per round in which they took a decision and $5$ points for each round left between the consensus round and the maximum round, plus $15$ points if the consensus was achieved in the innovation color (minority at the beginning).

\begin{figure}[H]
    \centering
    \includegraphics[width = \linewidth]{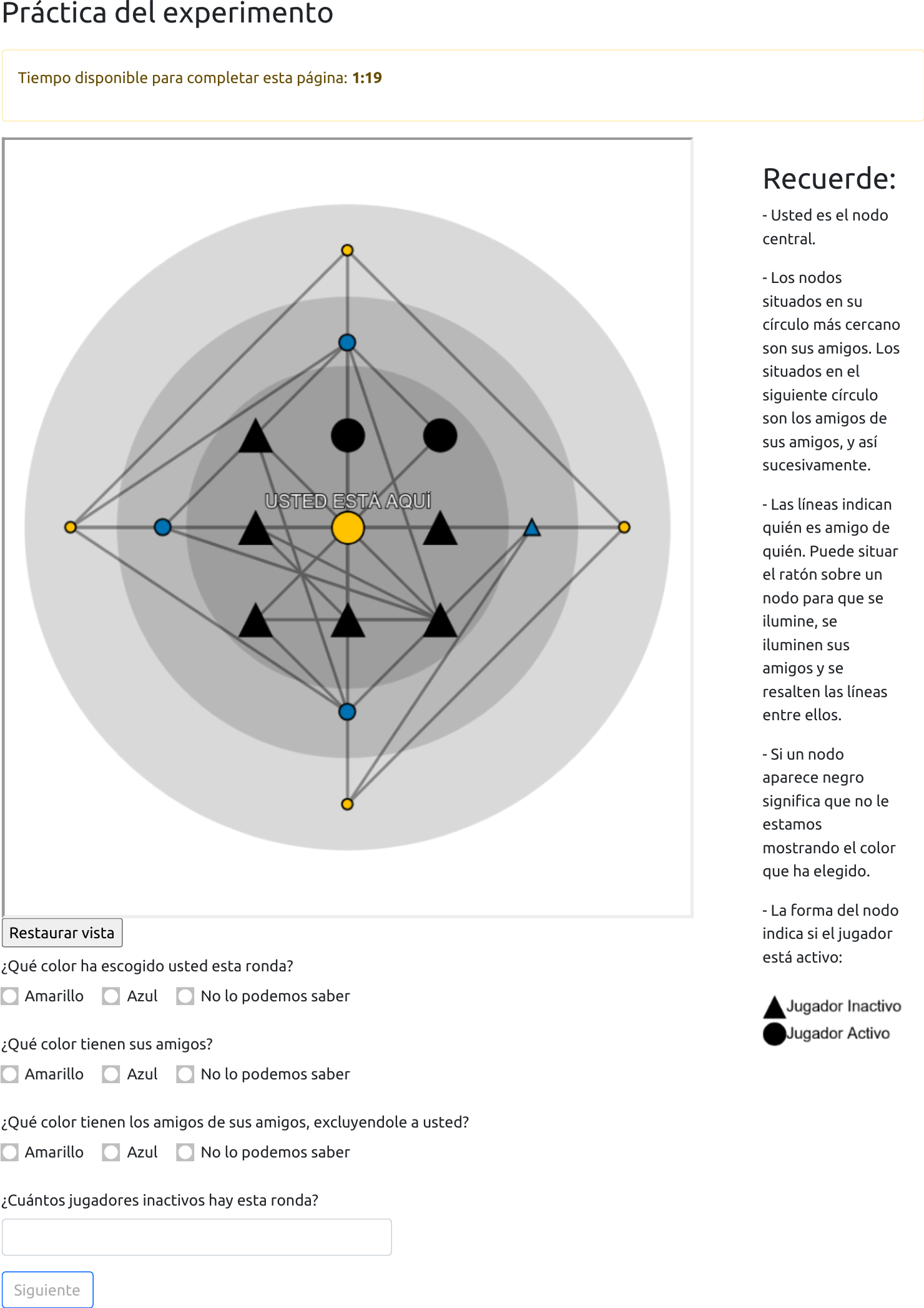}
    \caption{Practice network shown in the Instructions phase.}
    \label{fig:Practica}
\end{figure}

After these explanations, we showed them an image similar to the networks where we will show them the information (Fig. \ref{fig:Practica}). They are networks displayed in circular layers, such that the central node corresponds to the participant and the rest of nodes are disposed in further layers depending on the topological distance to the central one. To visually distinguish easily the distance of each layer, we plotted gray figures in the background which were darker as they were closer to the central node. Furthermore, a text indicating which node is the player is displayed on top of the central node. For each node in the network, two types of information are shown. Firstly, the color chosen by such node, being black if that information is not provided to the player. Secondly, the shape of the node is a circle if it represents an active player (who has made a decision in this round) or a triangle if the represented player is inactive.

\begin{figure}[H]
    \centering
    \includegraphics[width = \linewidth]{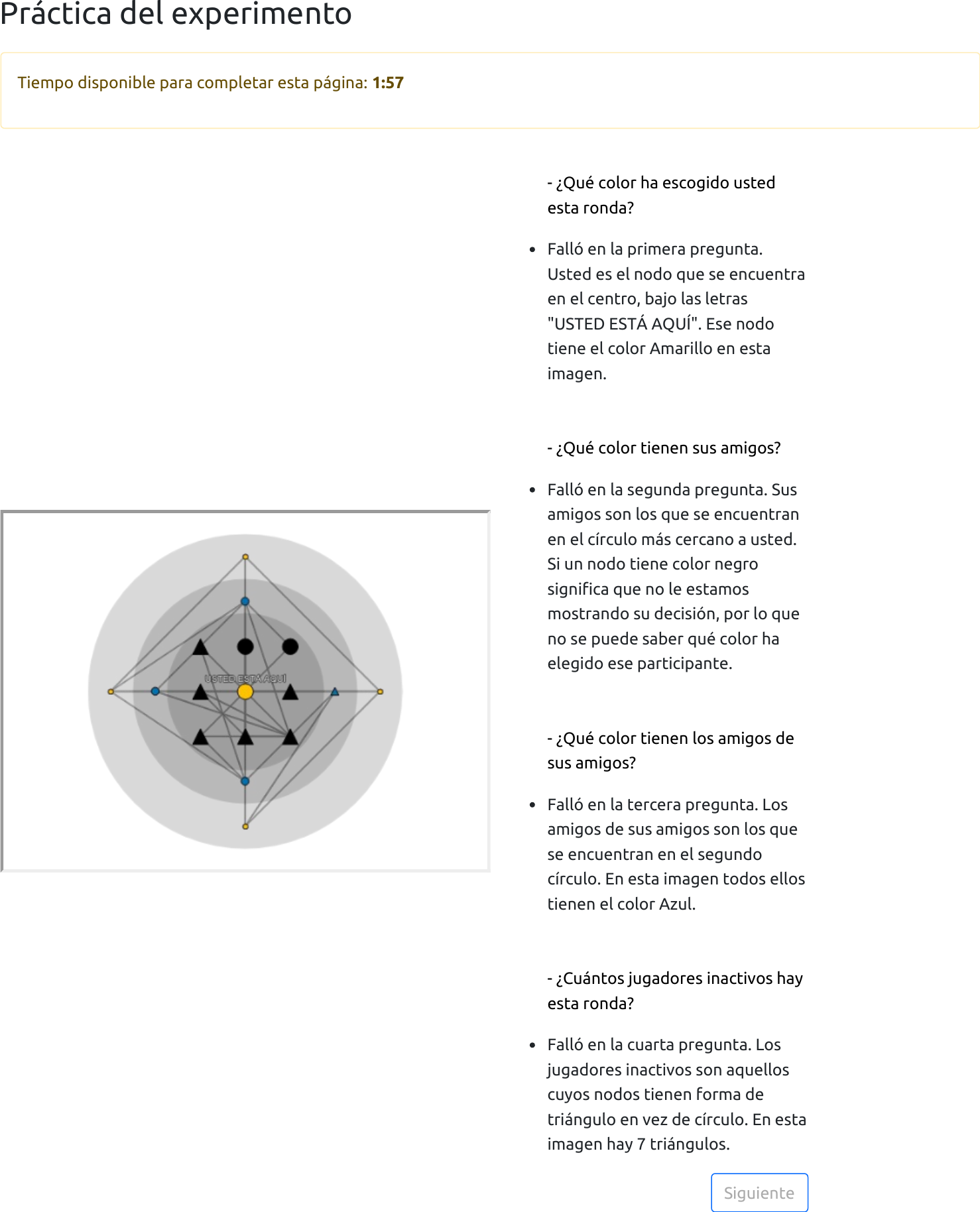}
    \caption{Explanation of the wrong answers to the questions in the Instructions phase.}
    \label{fig:InstructionsAnswers}
\end{figure}

To capture if instructions are clearly understood, 4 questions had to be answered by the participants: Which color have you chose this round?, Which color was chosen by your friends?, Which color was chosen by the friends or your friends, excluding yourself? and How many inactive players are there in this round?.  The first three questions had options "Blue", "Yellow" and "We can't know it", while the last one had to be answered by typing a number. After answering them, the next page showed if they answered correctly each question and, if they were wrong, they were given a brief explanation about which was the correct answer and why (Fig. \ref{fig:InstructionsAnswers}).

\begin{figure}[H]
    \centering
    \begin{subfigure}[b]{0.45\linewidth}
        \centering
        \includegraphics[width=\linewidth]{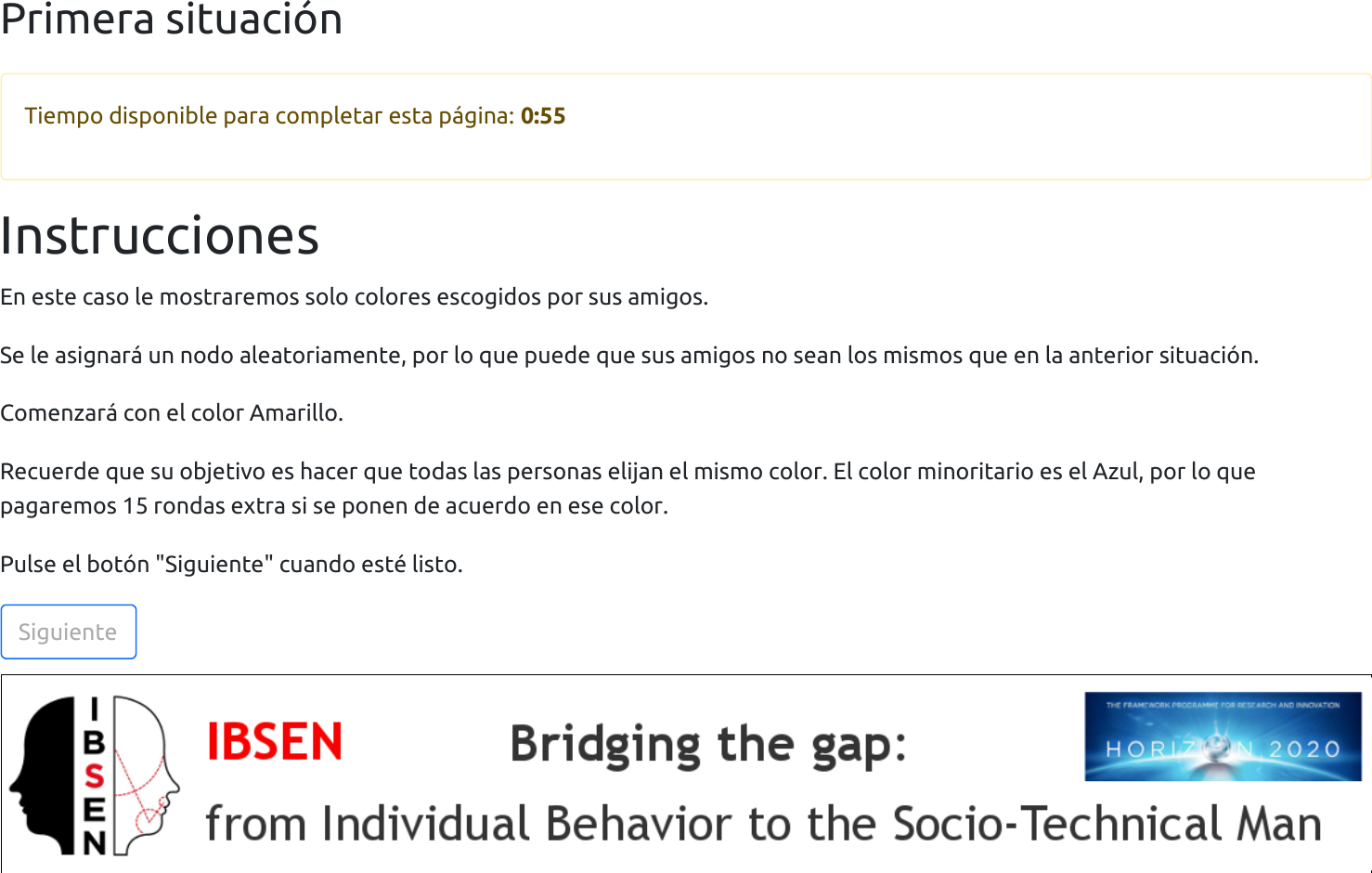}
        \caption{Explanation of the game for Setting I.}
        \label{fig:ExplanationAmAz}
    \end{subfigure}
    \hspace{0.01\linewidth}
    \begin{subfigure}[b]{0.45\linewidth}
        \centering
        \includegraphics[width=\linewidth]{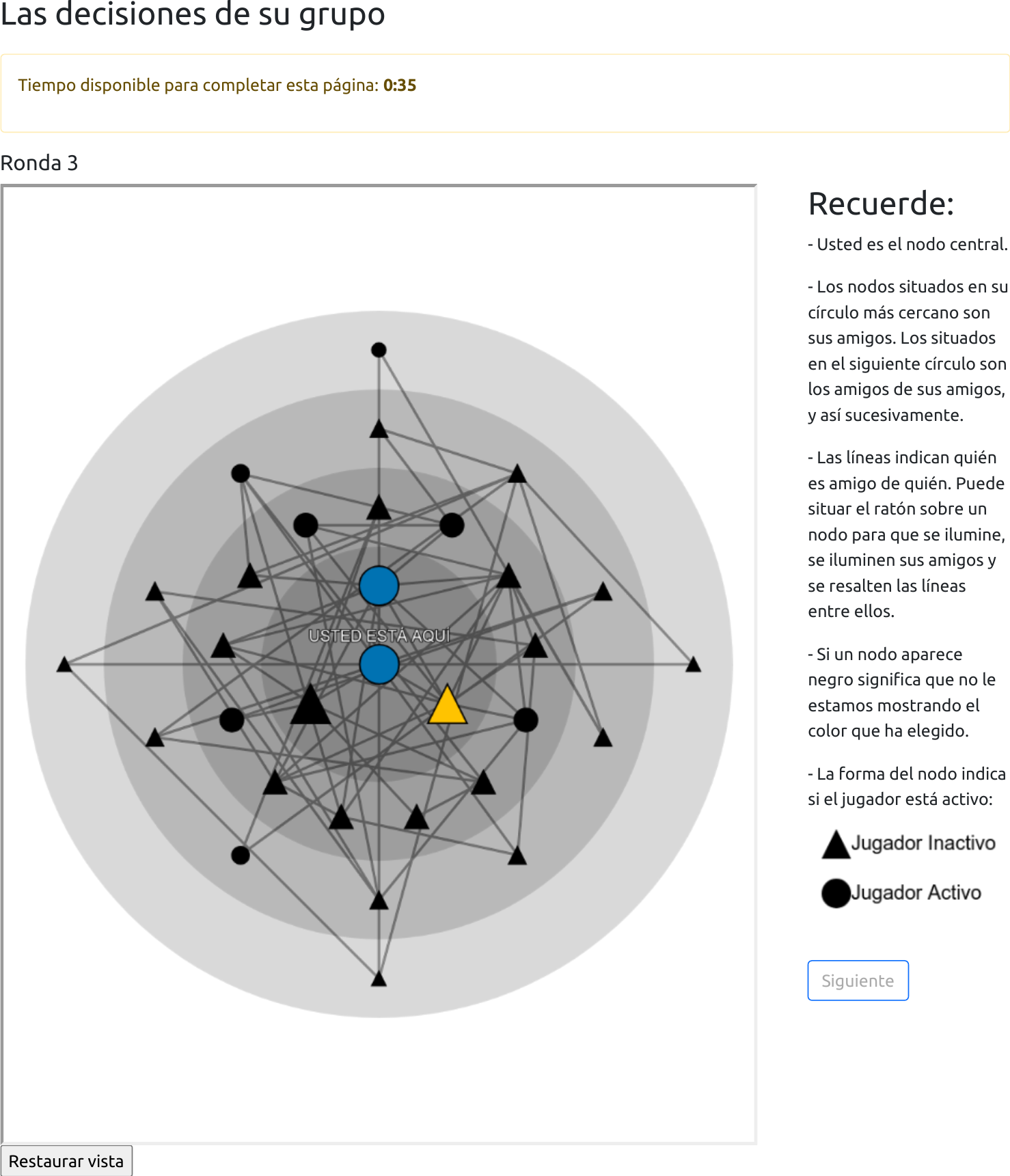}
        \caption{Resulting network page for Setting I.}
        \label{fig:NetworkAmAz}
    \end{subfigure}
    \\
    \vspace{0.5cm}
    \begin{subfigure}[b]{0.45\linewidth}
        \centering
        \includegraphics[width=\linewidth]{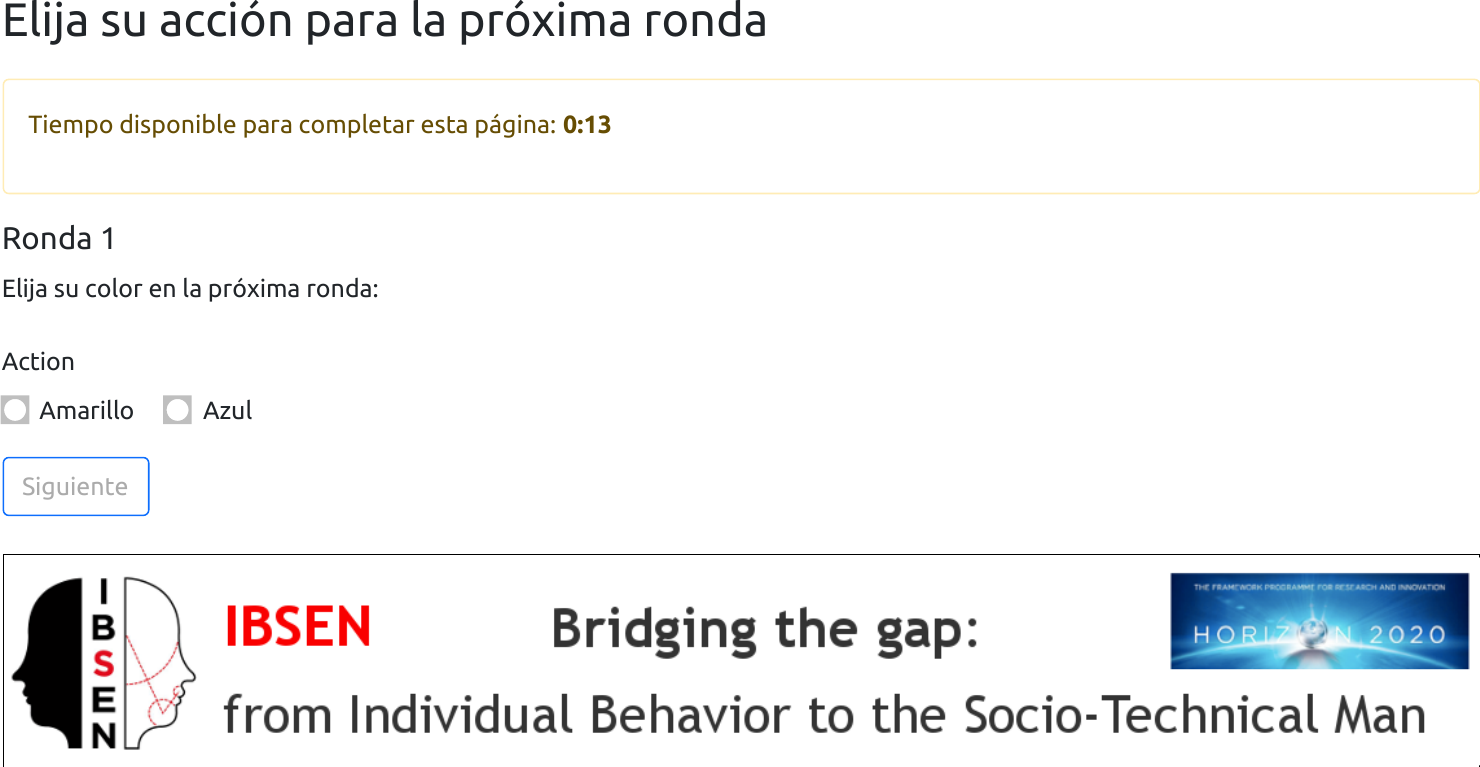}
        \caption{Choice page for Setting I.}
        \label{fig:ChoiceAmAz}
    \end{subfigure}
    \hspace{0.01\linewidth}
    \begin{subfigure}[b]{0.45\linewidth}
        \centering
        \includegraphics[width=\linewidth]{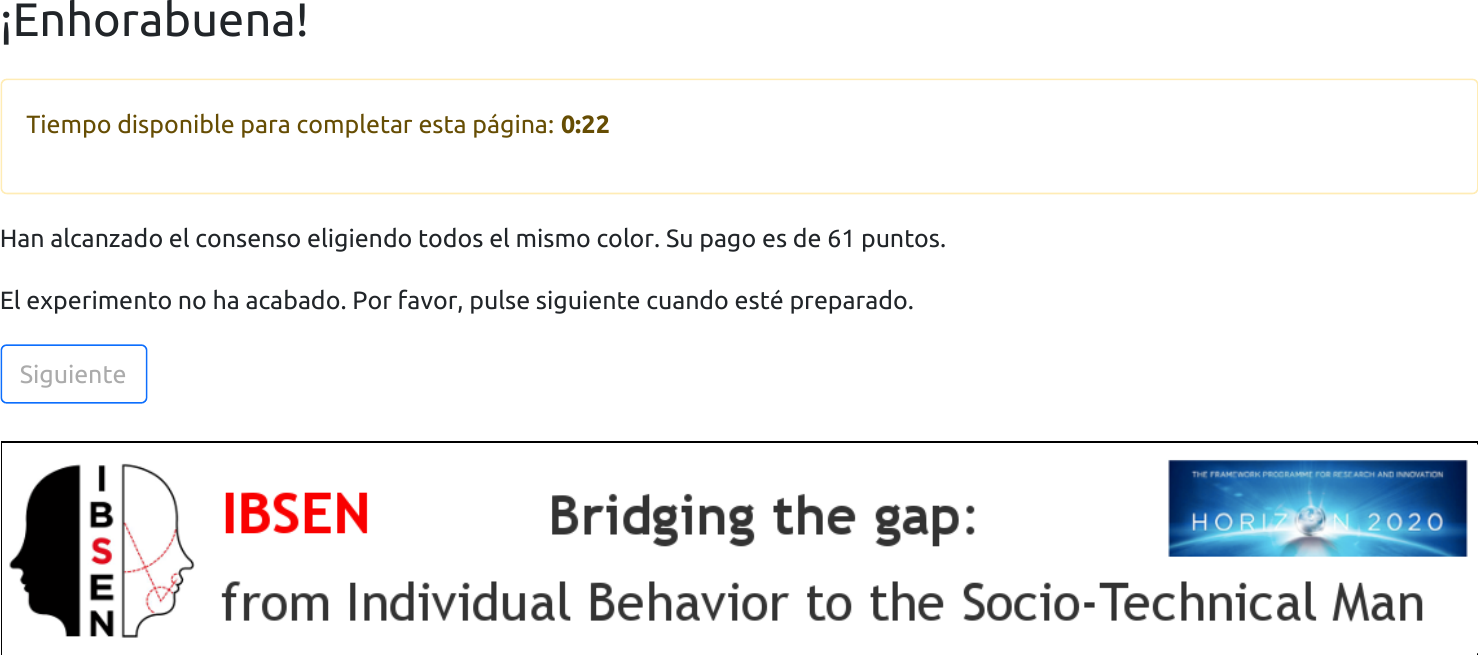}
        \caption{Results page for Setting I.}
        \label{fig:ResultsAmAz}
    \end{subfigure}
    \caption{Overview of the different pages for Setting I.}
\end{figure}

After all players completed the Instructions phase, the Setting I started. The first pages was a brief explanation of this game (Fig. \ref{fig:ExplanationAmAz}), indicating that in this phase there will be several rounds and in each of them we will show them two of their friends (nodes at distance 1). This friends are chosen randomly, excluding inactive players if possible. They were also told between which two colors they had to choose (Blue and Yellow), which of them was the initially minority color for which we would pay them more points (Blue) and which color was initially assigned to them. After that, successive pages of the resulting network (Fig. \ref{fig:NetworkAmAz}) and a choice page (Fig. \ref{fig:ChoiceAmAz}), in which the two colors were displayed in random order to each participant in each round. If the choice of all participants is the same or if the maximum number of rounds is achieved, a final page with the result and points earned is shown (Fig. \ref{fig:ResultsAmAz}).

\begin{figure}[H]
    \centering
    \begin{subfigure}[b]{0.45\linewidth}
        \centering
        \includegraphics[width=\linewidth]{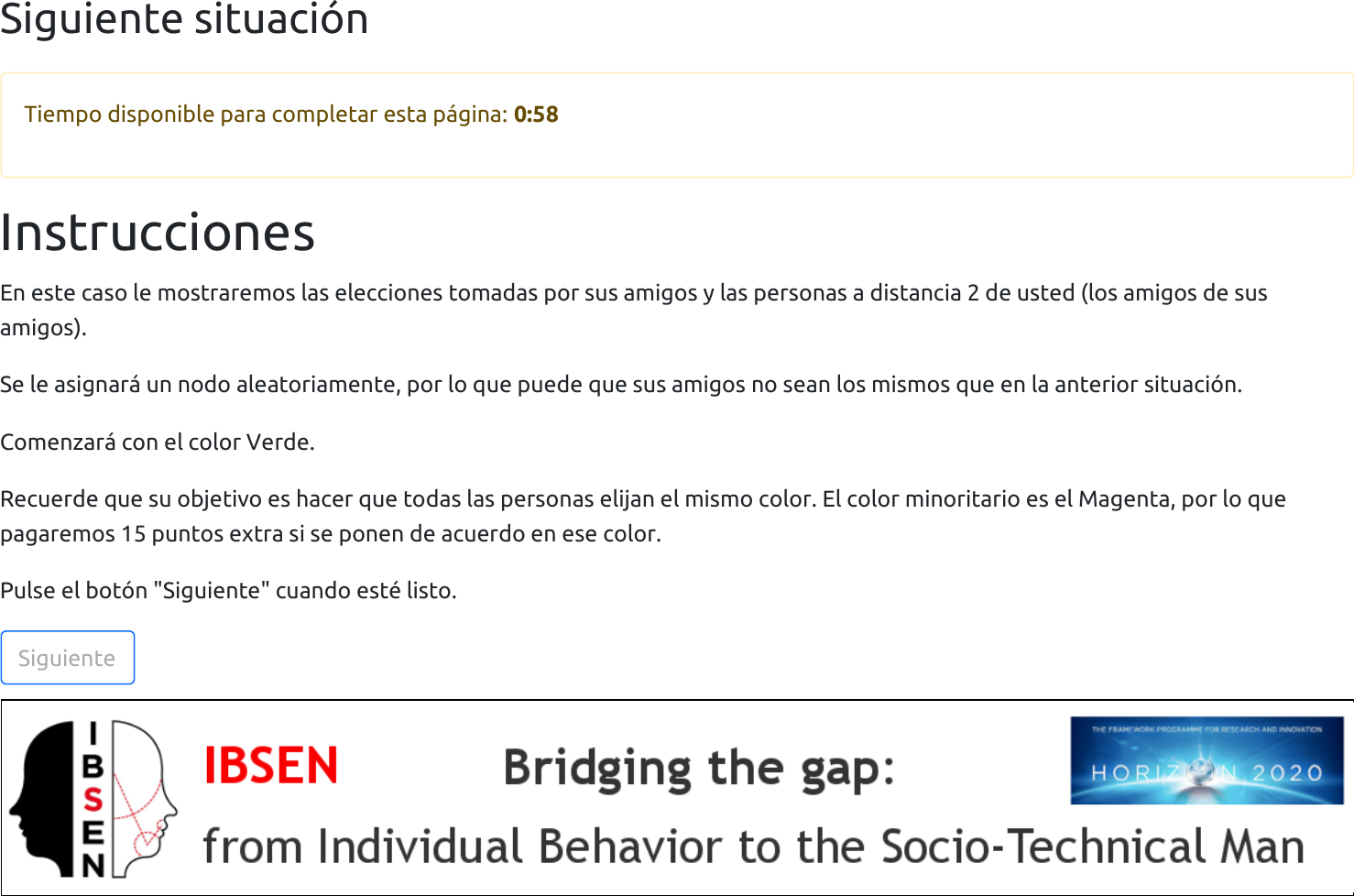}
        \caption{Explanation of the game for Setting II.}
        \label{fig:ExplanationVerMag}
    \end{subfigure}%
    \hspace{0.01\linewidth}
    \begin{subfigure}[b]{0.45\linewidth}
        \centering
        \includegraphics[width=\linewidth]{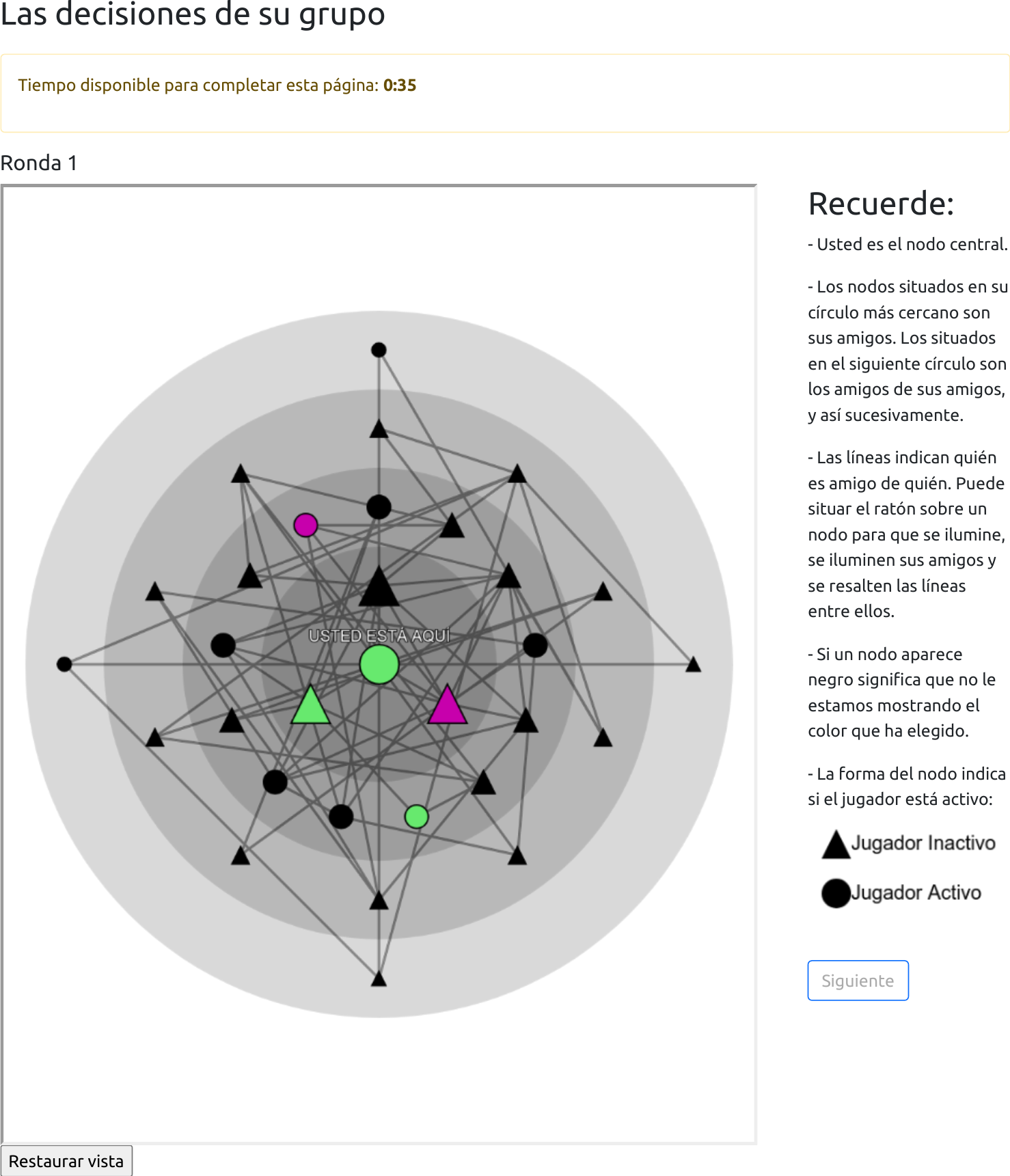}
        \caption{Resulting network shown for Setting II.}
        \label{fig:NetworkVerMag}
    \end{subfigure}
    \caption{Overview of explanation and network for Setting II.}
\end{figure}

\begin{figure}[H]
    \begin{subfigure}[b]{0.45\linewidth}
        \centering
        \includegraphics[width=\linewidth]{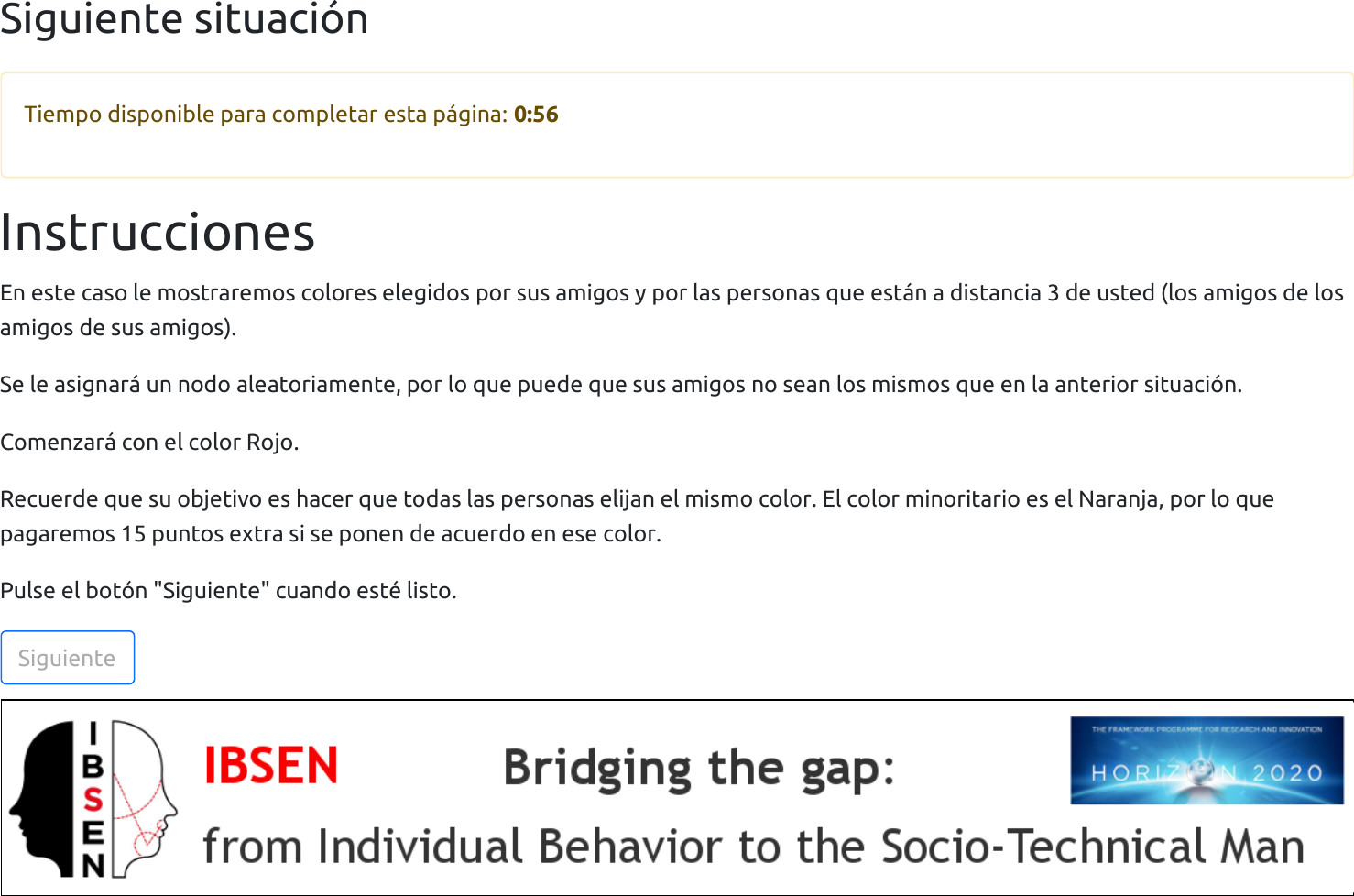}
        \caption{Explanation of the game for Setting III.}
        \label{fig:ExplanationRojNar}
    \end{subfigure}%
    \hspace{0.01\linewidth}
    \begin{subfigure}[b]{0.45\linewidth}
        \centering
        \includegraphics[width=\linewidth]{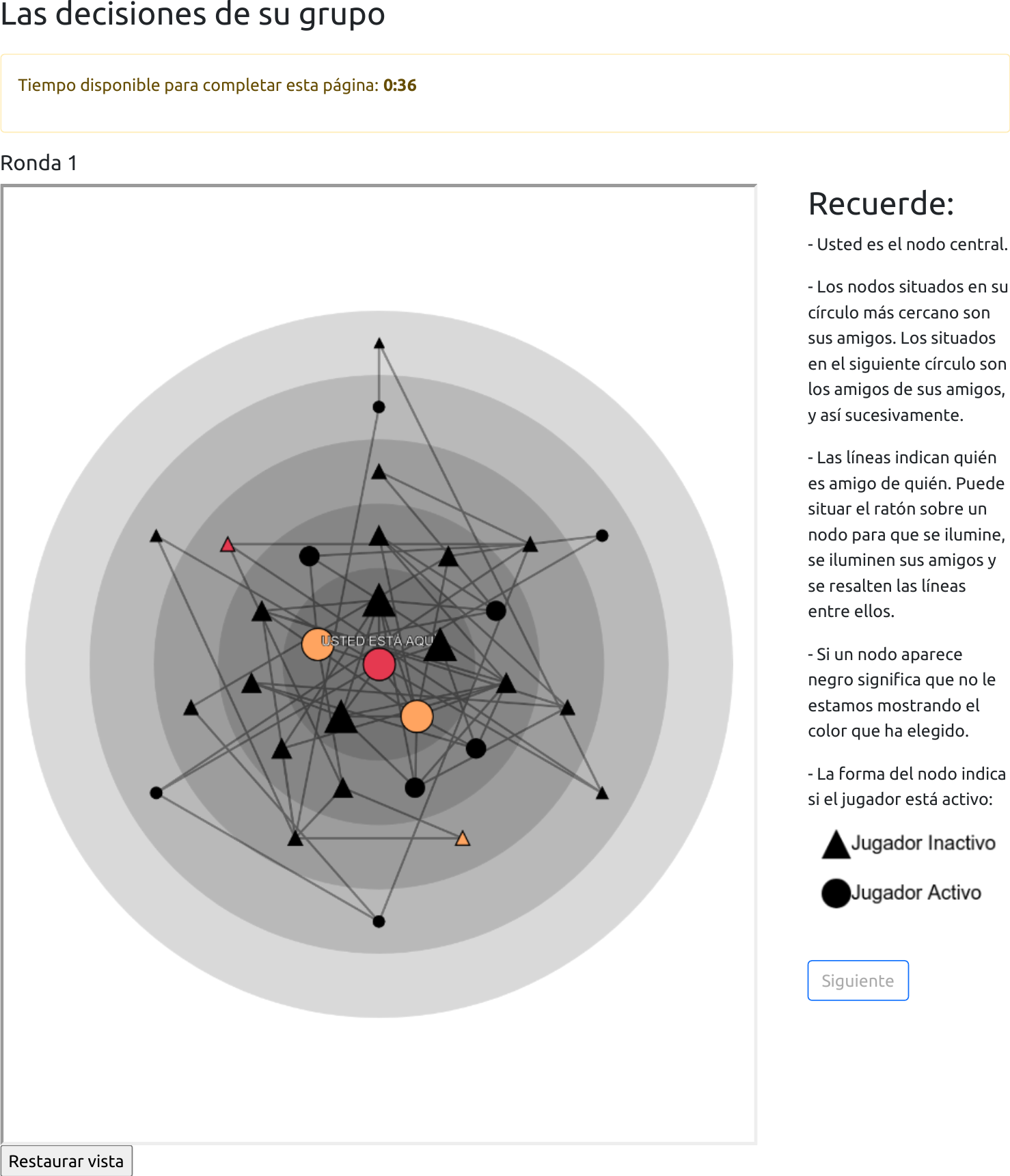}
        \caption{Resulting network shown for Setting III.}
        \label{fig:NetworkRojNar}
    \end{subfigure}
    \caption{Overview of explanation and network for Setting III.}
\end{figure}

\begin{figure}[H]
    \begin{subfigure}[b]{0.45\linewidth}
        \centering
        \includegraphics[width=\linewidth]{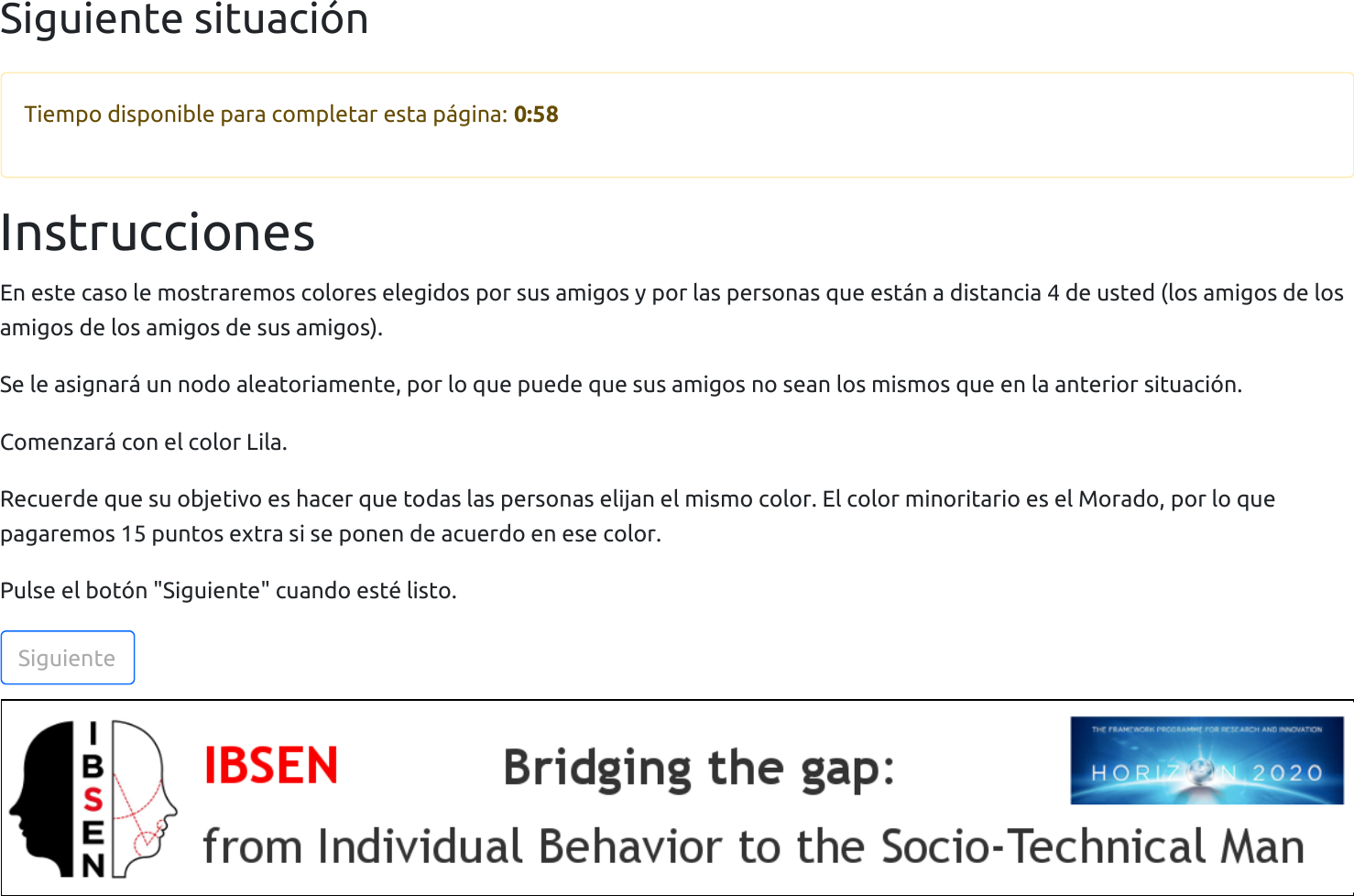}
        \caption{Explanation of the game for Setting IV.}
        \label{fig:ExplanationLilMor}
    \end{subfigure}%
    \hspace{0.01\linewidth}
    \begin{subfigure}[b]{0.45\linewidth}
        \centering
        \includegraphics[width=\linewidth]{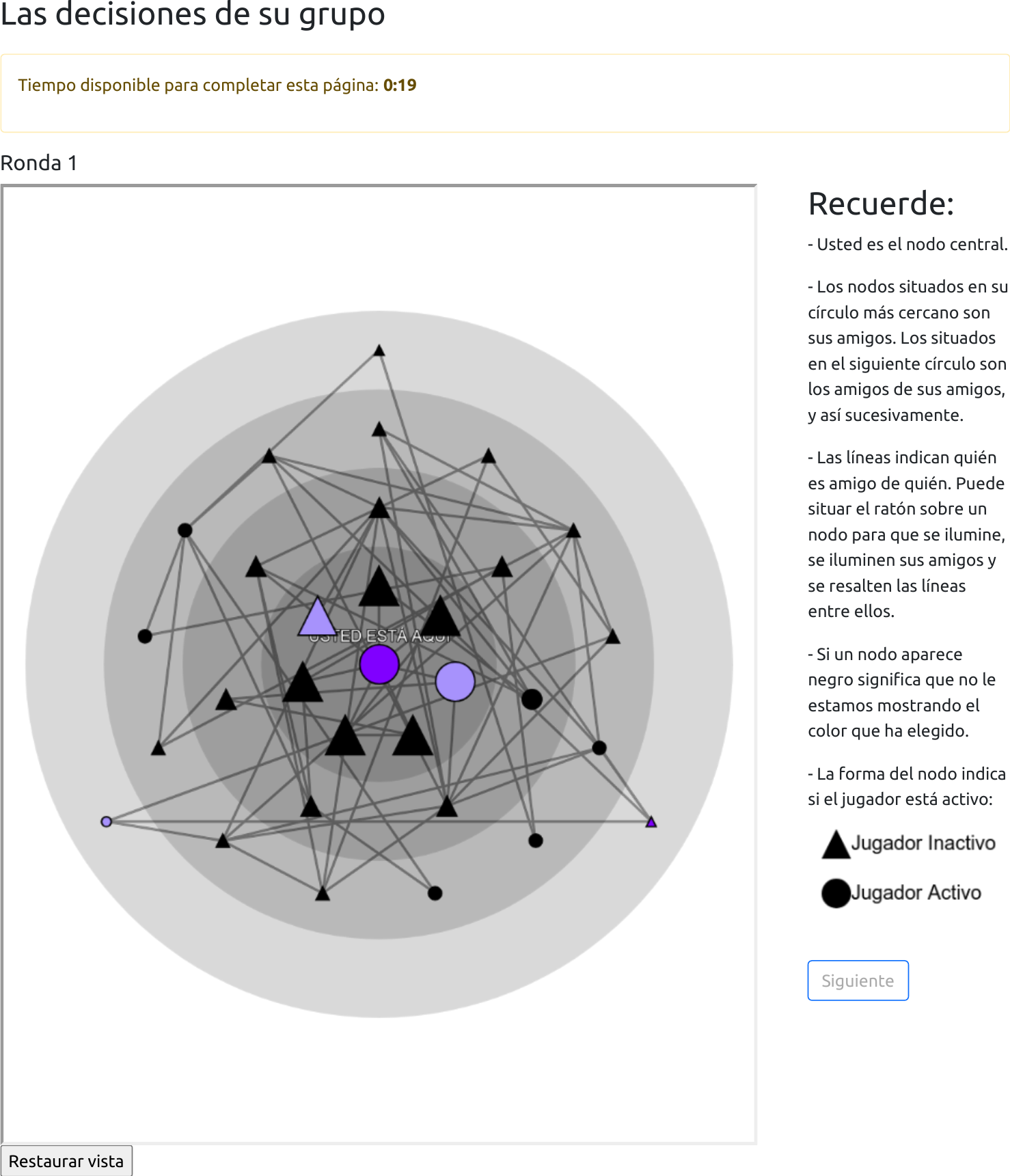}
        \caption{Resulting network shown for Setting IV.}
        \label{fig:NetworkLilMor}
    \end{subfigure}
    \caption{Overview of explanation and network for Setting IV.}
\end{figure}

Then, the three next phases consisted in Settings II, III and IV, differently ordered for each of the experiment sessions. In each of them, the same sequence of pages are shown, as for Setting I (an explanation, several rounds of the resulting network and the choice page and finally a result page with the points earned). In Setting II there were shown two random neighbors (at distance 1) and two random nodes at distance 2, always evading inactive players if possible. In this setting the colors to chose between are Green and Magenta, being the last one the minority color at the beginning. In Setting III there were shown also two random neighbors at distance 1 and two random nodes at distance 3, choosing between Orange (minority) and Red (majority). Finally, in Setting IV participants had to choose between Purple as minority color and Lilac, and we showed them two nodes at distance 1 and two nodes at distance 4. Explanation pages are shown in Figs. \ref{fig:ExplanationVerMag},\ref{fig:ExplanationRojNar},\ref{fig:ExplanationLilMor} and examples of the Network pages are in \ref{fig:NetworkVerMag},\ref{fig:NetworkRojNar},\ref{fig:NetworkLilMor}.

\begin{figure}[H]
    \begin{subfigure}[b]{0.45\linewidth}
        \centering
        \includegraphics[width = \linewidth]{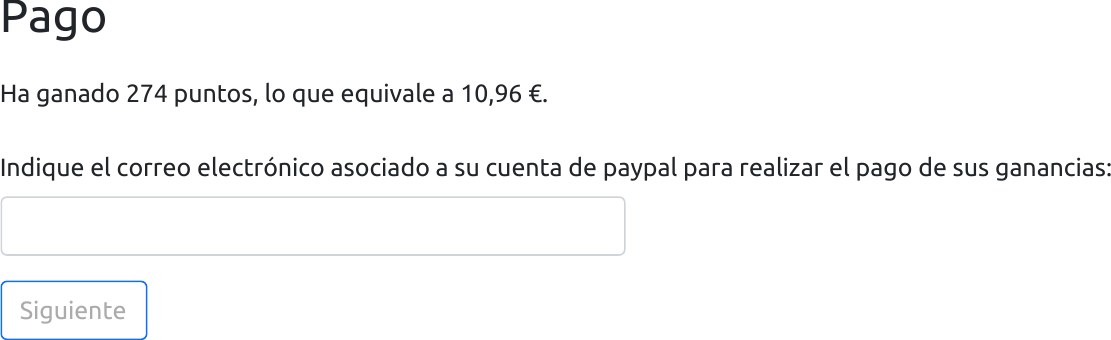}
        \caption{Payment page.}
        \label{fig:Payment}
    \end{subfigure}%
    \hspace{0.01\linewidth}
    \begin{subfigure}[b]{0.45\linewidth}
        \centering
        \includegraphics[width = \linewidth]{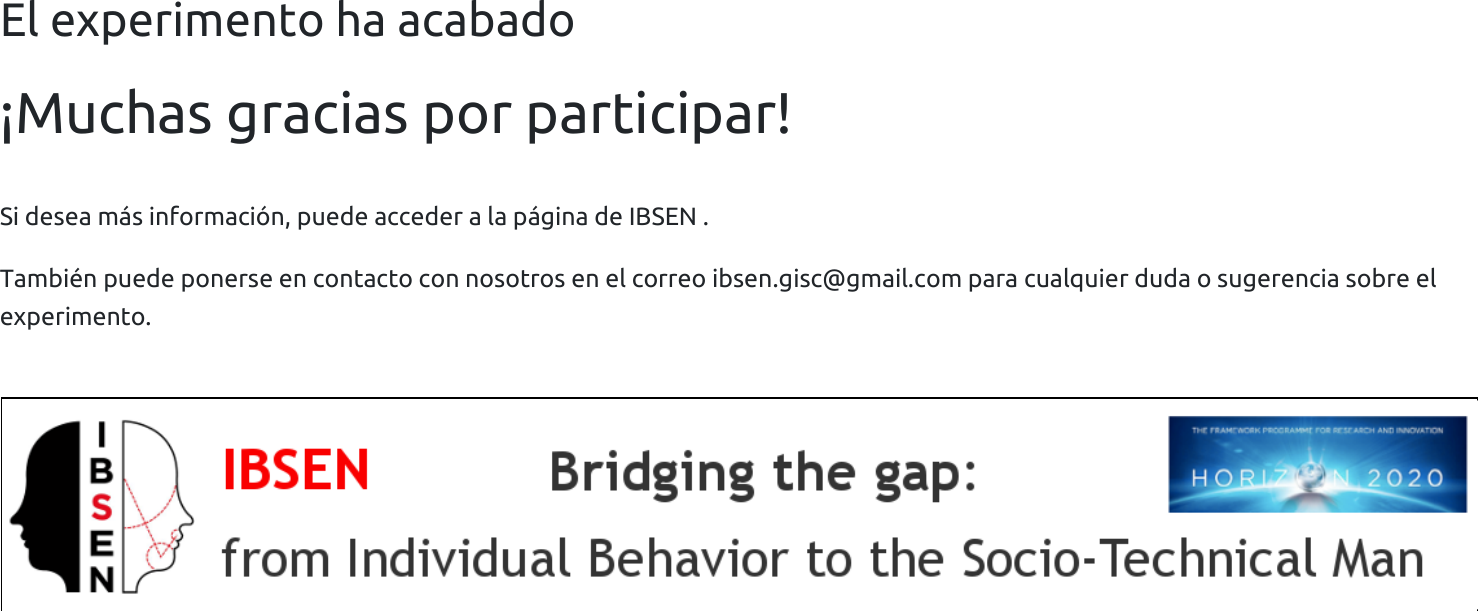}
        \caption{End-page.}
        \label{fig:End}
    \end{subfigure}
    \caption{Payment and end-page.}
\end{figure}

Finally, the last pages shown to the participants (Figs. \ref{fig:Payment},\ref{fig:End}) gave each participant the total sum of points they earned, its equivalence in euros and asked them to introduce their Paypal account. After that, a greeting final page was shown.

{\eject \pdfpagewidth=215mm \pdfpageheight=690mm \headheight=7pt
\begin{figure}[H]
    \centering
    \includegraphics[scale = 0.5]{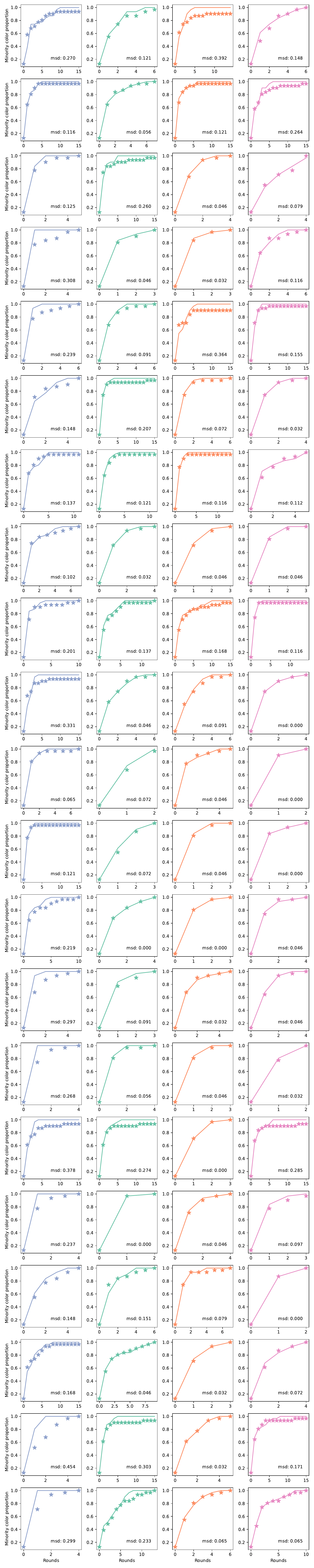}
    \caption{Individual results and fitted curves for each session and Setting of the experiment.}
\end{figure}
\pagestyle{empty} 

}

\section{Model adjustment with the median}
In the Experimental Results section, we assumed that, although our data does not look normally distributed, it was due to the small size effect, and the real distribution would be Normal ones. Nonetheless, we could perform a different approach without considering this normality assumption. For that purpose, instead of the mean and standard deviation, we should use the medians ($c_2 = 0.68$, $c_3 = 0.1$, $c_4 = 0.64$). As we did in the article with the t-test, we can compare the distributions using the Mann-Whitney U test. The $p$-values obtained are: $p\text{-value}(c_2,c_3) = 0.0524$, $p\text{-value}(c_2,c_4) = 0.1694$ and $p\text{-value}(c_3,c_4) = 0.87$. As well as it occurs with the normality assumptions and the t-test, we cannot statistically ensure that medians are not equal pairwise, but the $p$-value for the distributions of $c_2$ and $c_3$ is close to the standard limit value of $0.05$, so it may be considered valid due to the small sample size.

\begin{figure}[H]
    \centering
    \includegraphics[width = 0.5\linewidth]{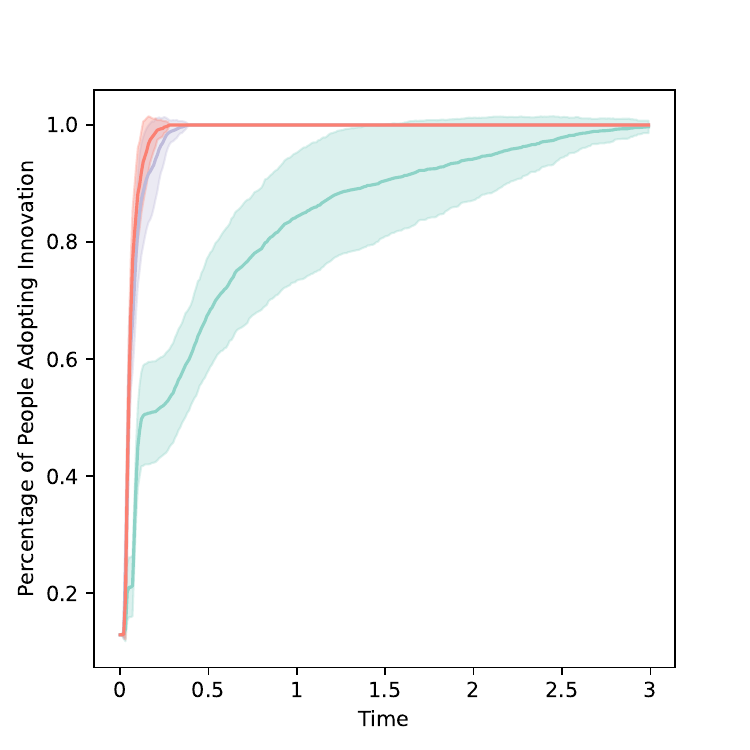}
    \caption{Mean plot of different diffusion models using $1000$ random initial conditions. Green line represents the classical (short distance) model, gray line represents the long-range model with the median parameters and red line represents the same model but with the mean parameters.}
    \label{fig:medianplot}
\end{figure}

Thus, we discard Setting IV as the less conclusive one from our data and use Settings II and III to do our median prediction model: $\dot{u}(t) \approx -(\mathcal{L}_1 + 0.68 \mathcal{L}_2 + 0.1 \mathcal{L}_3) u(t)$. In Fig.\ref{fig:medianplot} we plot the mean behavior of the classical model (using only $\mathcal{L}_1$), our model with the mean parameters used in the main text and our model with the median values presented here, randomizing the initial conditions. We can observe that the change between these two long-range models is minimum when picking the mean or the median values.

\section{Other coefficient decays for long-range interactions}

For the $d$-path Laplacian operators, there has been two types of transformations that have been studied \cite{estrada-2017,estrada-2018}: Mellin transformation, that corresponds to coefficients $c_d = d^{-s}$ for $s>0$, and Laplace transformation, whose coefficients are given by $c_d = e^{-d \lambda}$ for $\lambda > 0$. The first one has been more deeply investigated as it was proven that, for certain values of the parameter $s$, it models a superdiffusive process in a $1$-dimensional infinite path graph or in a $2$-dimensional infinite square lattice.

\begin{figure}[H]
    \centering
    \includegraphics[width = \linewidth]{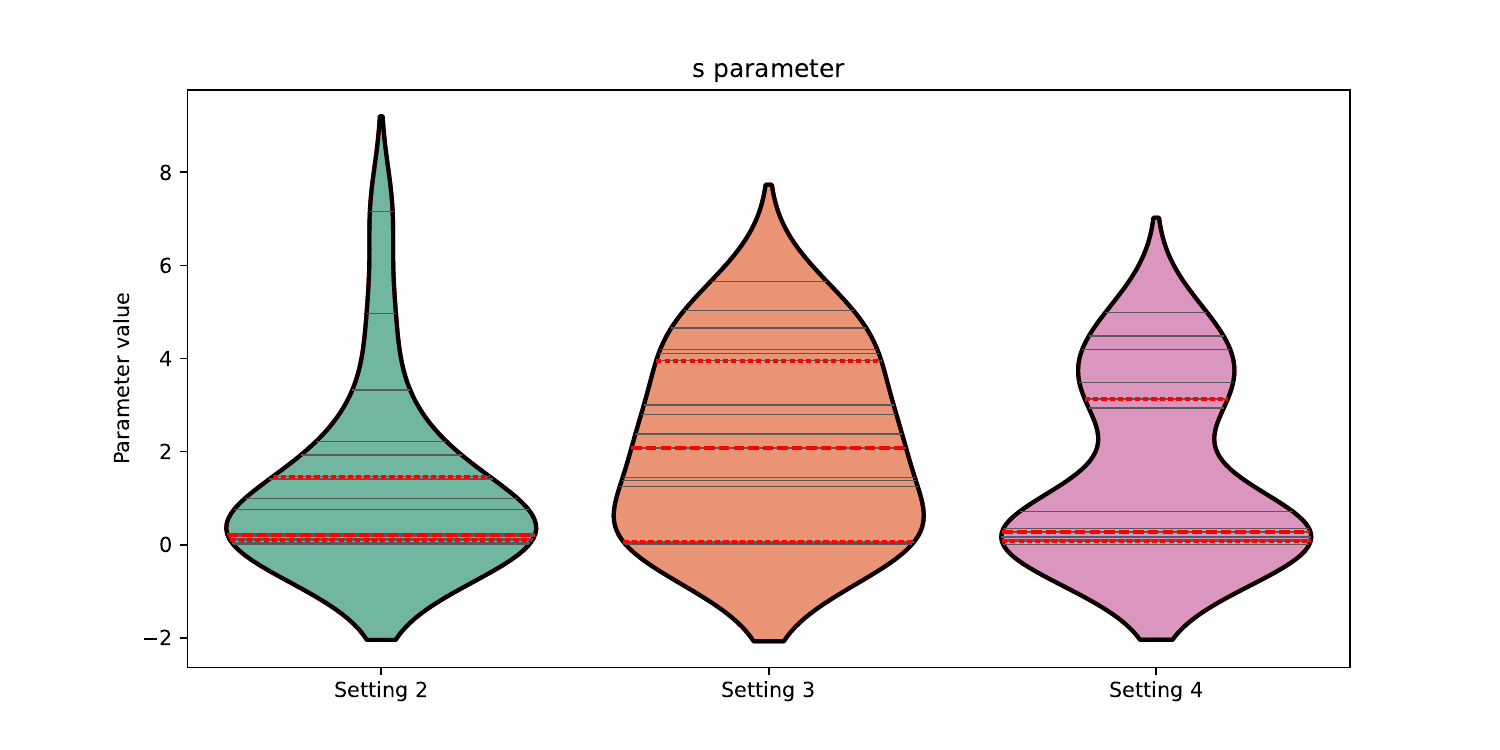}
    \caption{Distribution of the values of the $s$ parameters for each of the sessions and Settings.}
    \label{fig:sParam}
\end{figure}

\begin{figure}[H]
    \centering
    \includegraphics[width = \linewidth]{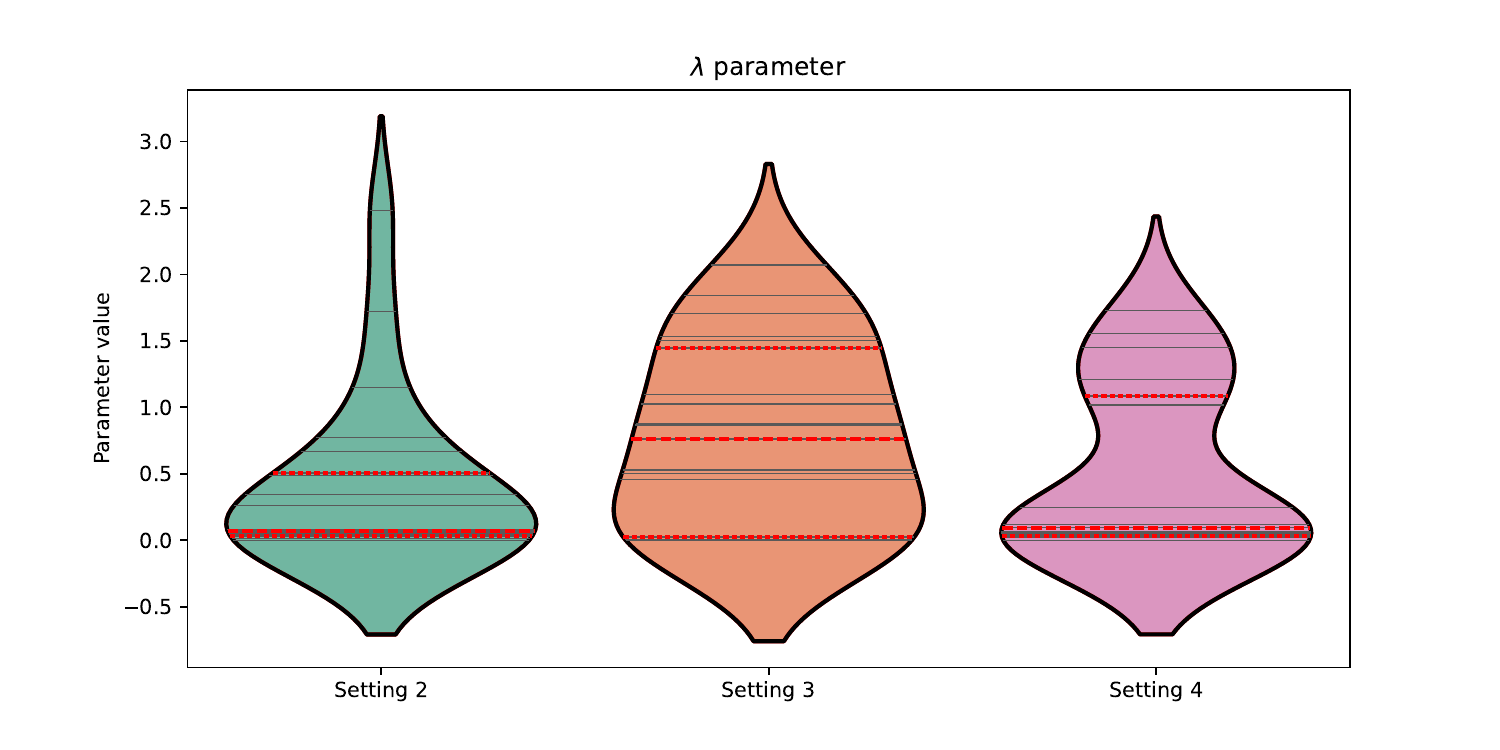}
    \caption{Distribution of the values of the $\lambda$ parameters for each of the sessions and Settings.}
    \label{fig:lambdaParam}
\end{figure}

We calculated which values of the parameters $s$ and $\lambda$ would correspond to each of the experiments and Settings within them (Figs. \ref{fig:sParam} and \ref{fig:lambdaParam}).

Focusing on the $s$ parameters, we obtain that, in both cases, the median of Setting II and IV are similar ($s = 0.21$ and $s = 0.27$, respectively), while for Setting III it increases to $s = 2.08$. This reflects the fact that in Setting III there could be spotted two different groups of values for $c_3$, one with values close to $1$ and other with lower values, between $0$ and $1/3$. It is the first of them that pushes the median value higher, which means a curve slowly decaying (further nodes are still considered as important).

\begin{figure}[H]
    \centering
    \includegraphics[width = 0.6\linewidth]{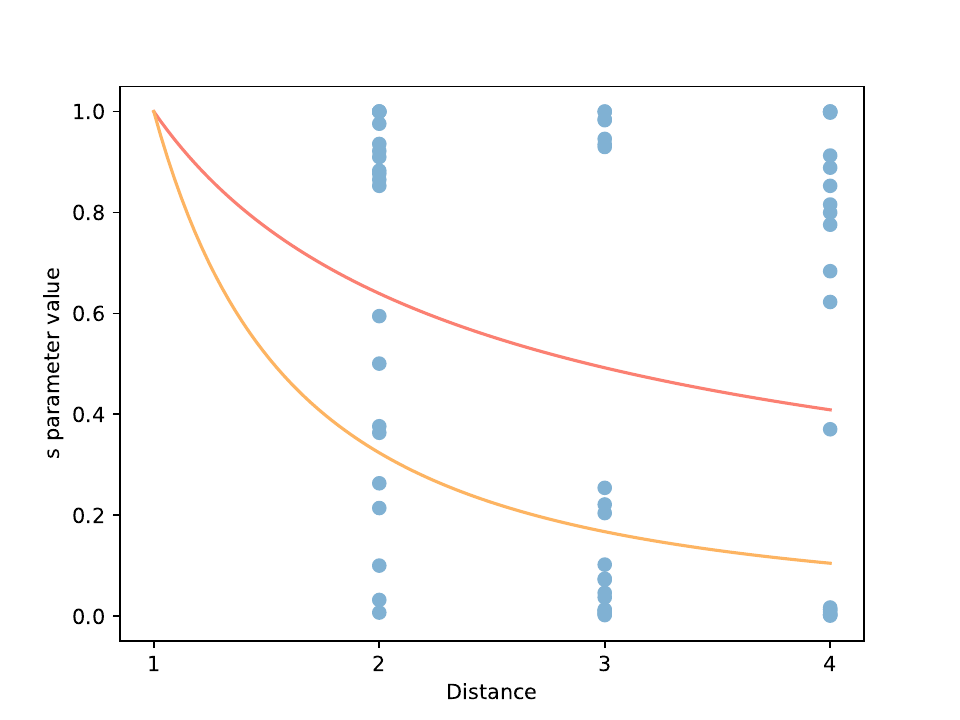}
    \caption{Power-law fitting of the data (blue dots) by MSE (red line) and by the mean of the transformed $s$ parameters (orange line).}
    \label{fig:DifferenceFit}
\end{figure}

Another relevant note when considering such functions is to not group by Settings for fitting. This is, if we calculated the mean obtained parameter for each of the sessions, the best parameter to fit that data would be the mean of the values obtained. For our experiments, this would give a value $s = 1.63 > 1$. On the other hand, we could fit the function parameter to the brute data $c_d$, which in our case gives $s = 0.65 < 1$. This difference (see Fig. \ref{fig:DifferenceFit}), result of the logarithmic transformations needed to go from $c_d$ to the $s$ parameters, shows that, depending on the contexts (Settings), the same group of people may consider peer influence with different rules (decay functions). This should be considered in future researches replicating or expanding these experiments.

\section{Demographics}
We recruited 596 participants in 21 sessions completed along 4 months, plus $3$ testing sessions and $2$ more canceled ones. The number of active participants in total was $596$, who received a mean payment of $4,50$€. 63.8\% of the were female, 35. 9\% were male and 0. 3\% were non-binary.  The average age was 30.40 (sd = 11.22). The age distribution of the participants per gender is shown in the following table and figures.

\begin{table}[h!]
\centering
\caption{Gender distribution per experimental session}
\label{table:gender}
\resizebox{\linewidth}{!}{\begin{tabular}{lllllllllllllllllllllll}
\toprule
\multirow{2}{*}{Gender} & \multirow{2}{*}{Total} & \multicolumn{21}{c}{Experimental session} \\
\cmidrule{3-23}
& & 1 & 2 & 3 & 4 & 5 & 6 & 7 & 8 & 9 & 10 & 11 & 12 & 13 & 14 & 15 & 16 & 17 & 18 & 19 & 20 & 21 \\
\midrule
Female     & 380   & 14           & 18           & 17           & 17           & 19           & 15           & 19           & 21           & 22           & 22            & 21            & 12            & 19            & 18            & 19            & 17            & 19            & 18            & 16            & 16            & 21            \\
Male       & 214   & 12           & 8            & 13           & 11           & 9            & 13           & 8            & 6            & 9            & 8             & 10            & 9             & 12            & 11            & 12            & 13            & 11            & 8             & 14            & 10            & 7             \\
Non-binary & 2     &              &              &              &              &              &              &              &              &              & 1             &               &               &               &               &               &               &               & 1             &               &                 & \\  
\bottomrule              
\end{tabular}}
\end{table}

\begin{figure}[H]
\centering
\includegraphics[width=\textwidth]{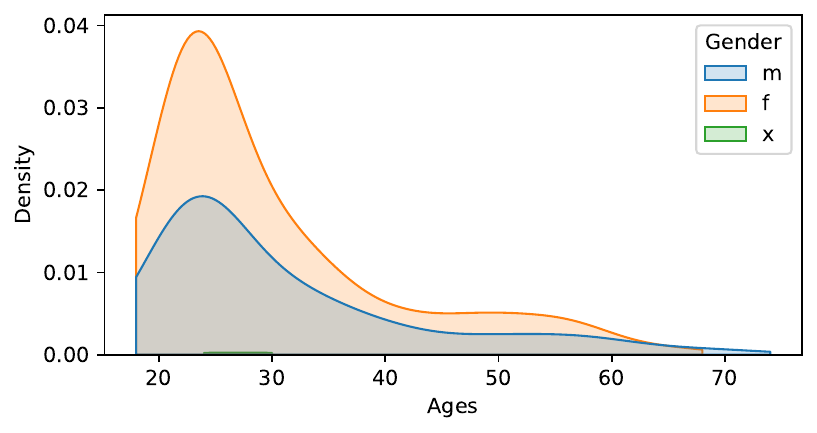}
\caption{Probability density function (PDF) of participant ages, per gender (m: male, f: female, x: non-binary.  PDFs per gender are scaled by the number of observations such that the total area under all densities sums up to 1.}
\label{fig:genderdistrib}
\end{figure}

\begin{figure}[H]
\centering
\includegraphics[width=0.87\textwidth]{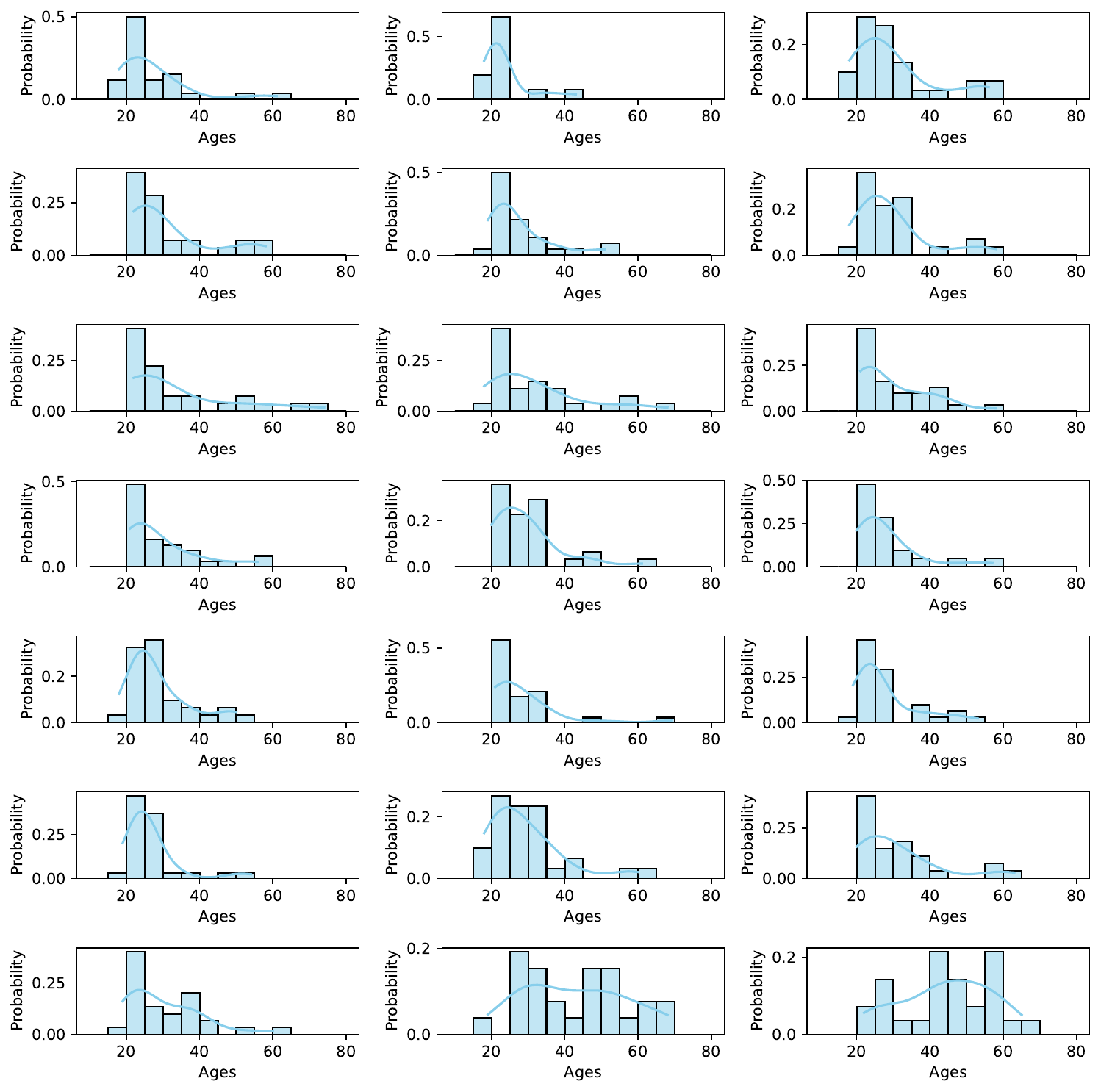}
\caption{Age probability function per experimental session. From top-left (session 1) to bottom-right (session 21). Bins width is five years.}
\label{fig:agedistrib}
\end{figure}